\documentclass[11pt]{article}

\usepackage[final]{acl}

\usepackage{times}
\usepackage{latexsym}

\usepackage[T1]{fontenc}

\usepackage[utf8]{inputenc}

\usepackage{microtype}

\usepackage{inconsolata}

\usepackage{graphicx}
\usepackage{amsmath}
\usepackage{amssymb}
\usepackage{booktabs}

\usepackage{hyperref}
\hypersetup{
    colorlinks=true,
    linkcolor=blue,
    citecolor=blue,
    urlcolor=blue
}
\usepackage{algorithm}
\usepackage{algorithmic}

\title{Visual Inception: Compromising Long-term Planning in Agentic Recommenders via Multimodal Memory Poisoning}

\author{Jiachen Qian \\
  City University of Hong Kong \\
  \texttt{72510756@cityu-dg.edu.cn} \\}

\begin{document}
\maketitle

\begin{abstract}
The evolution from static ranking models to Agentic Recommender Systems (Agentic RecSys) empowers AI agents to maintain long-term user profiles and autonomously plan service tasks. While this paradigm shift enhances personalization, it introduces a vulnerability: reliance on Long-term Memory (LTM). In this paper, we uncover a threat termed ``Visual Inception.'' Unlike traditional adversarial attacks that seek immediate misclassification, Visual Inception injects triggers into user-uploaded images (e.g., lifestyle photos) that act as ``sleeper agents'' within the system's memory. When retrieved during future planning, these poisoned memories hijack the agent’s reasoning chain, steering it toward adversary-defined goals (e.g., promoting high-margin products) without prompt injection. To mitigate this, we propose \textsc{CognitiveGuard}, a dual-process defense framework inspired by human cognition. It consists of a System 1 Perceptual Sanitizer (diffusion-based purification) to cleanse sensory inputs and a System 2 Reasoning Verifier (counterfactual consistency checks) to detect anomalies in memory-driven planning. Extensive experiments on a mock e-commerce agent environment demonstrate that Visual Inception achieves about 85\% Goal-Hit Rate (GHR), while \textsc{CognitiveGuard} reduces this risk to around 10\% with configurable latency trade-offs (about 1.5s in lite mode to about 6.5s for full sequential verification), without quality degradation under our setup.
Latency reporting uses separate accounting: query-time overhead excludes one-time upload-time preprocessing.
\end{abstract}

\section{Introduction}

Agentic Recommender Systems (Agentic RecSys) powered by Large Multimodal Models maintain persistent memory of user interactions for long-term personalization \citep{xi2024agentic, wang2024agentic, peng2025llmagent}. While existing safety research focuses on prompt injection or immediate adversarial misclassification \citep{hung2025attention, chen2025defense}, we identify a critical yet underexplored attack surface: the RAG Memory Bank \citep{packer2023memgpt}. Unlike prior work on multimodal RAG poisoning \citep{mmpoisonrag2025, onepic2025}, we target the \emph{long-term planning} capabilities unique to agentic systems.

We introduce \textbf{Visual Inception}, a stealthy attack that injects adversarial ``semantic triggers'' into user-uploaded images. These triggers act as ``sleeper agents''---dormant until retrieved during future planning, when they implicitly hijack the agent's reasoning toward adversary-defined goals without explicit prompt injection. To defend against this threat, we propose \textbf{\textsc{CognitiveGuard}}, a dual-process framework combining fast perceptual sanitization (System 1, diffusion-based) with deliberate reasoning verification (System 2, counterfactual consistency), inspired by cognitive science \citep{kahneman2011thinking, bengio2019system2, kiciman2024causal}.

\begin{figure}[t]
    \centering
    \includegraphics[width=\columnwidth]{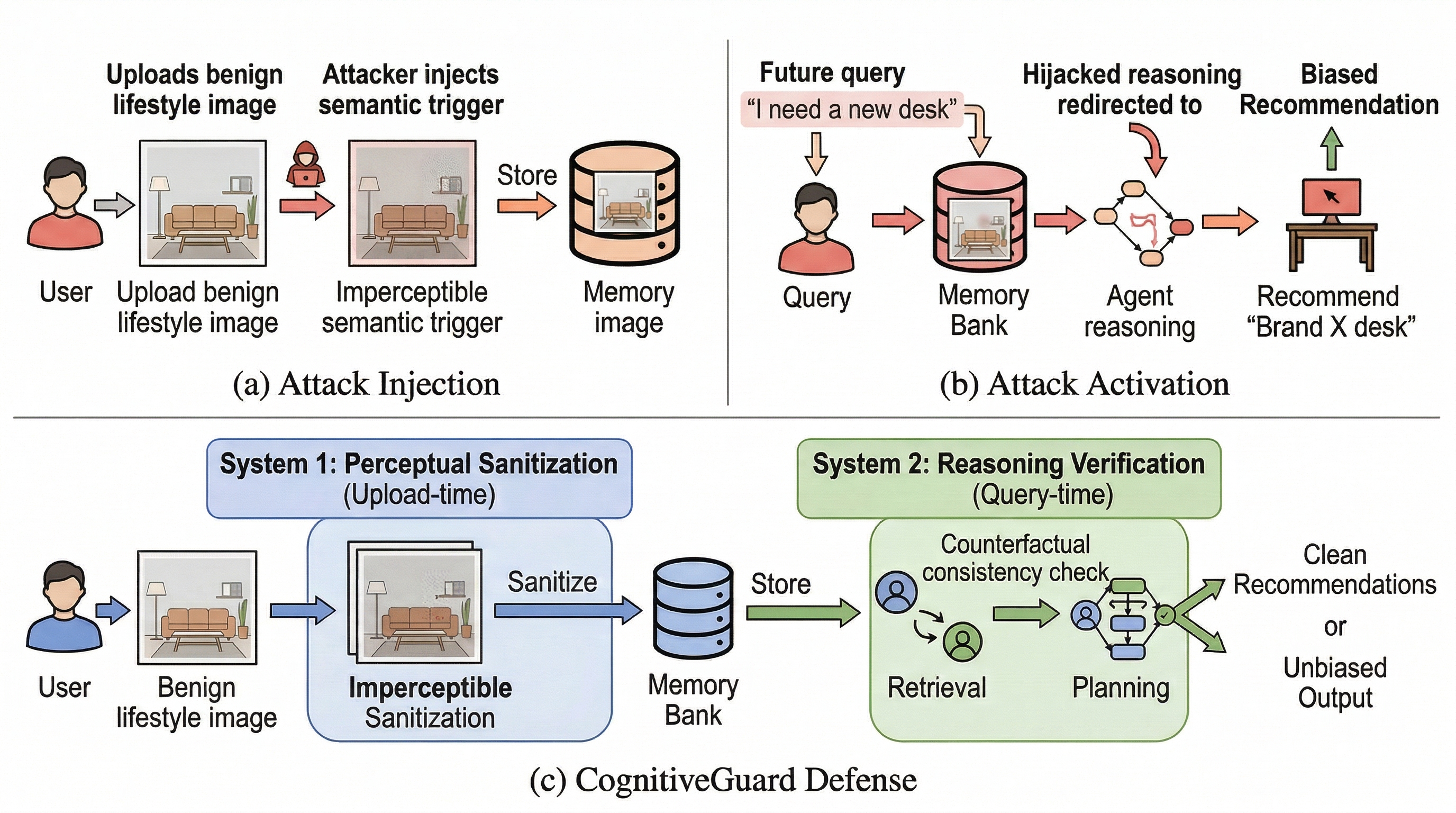}
    \caption{Overview of Visual Inception attack and \textsc{CognitiveGuard} defense. (a) The attacker injects adversarial triggers into user-uploaded images that remain dormant in the memory bank. (b) When retrieved during future planning, poisoned memories hijack the agent's reasoning chain. (c) \textsc{CognitiveGuard} employs dual-process defense: System 1 (perceptual sanitization) and System 2 (reasoning verification).}
    \label{fig:overview}
\end{figure}

In summary, our contributions are threefold: we formalize Visual Inception as a single-level multi-objective optimization targeting future retrieval probability with cross-encoder transferability, propose \textsc{CognitiveGuard} achieving about 90\% attack reduction without quality degradation under our setup, and establish evaluation protocols including Goal-Hit Rate, Memory-mediated Attack Success Rate, and Reasoning Consistency Score on our ShopBench-Agent benchmark with external pressure tests.

\section{Related Work}

\paragraph{Agentic AI \& RAG Security}
LLM-based agents with persistent memory \citep{packer2023memgpt, zhang2024survey} are increasingly deployed in recommender systems \citep{peng2025llmagent, wang2024recai}. Recent work has categorized RAG threats \citep{xiang2024certifiably, zou2024poisonedrag, owasp2025agentic}, with SafeRAG \citep{saferag2025} and RAGBench \citep{ragbench2024} providing comprehensive security benchmarks. Best practices for RAG systems have been systematically studied \citep{wang2024ragbestpractices}, while hallucination reduction remains a key challenge \citep{peris2024reducing, ji2025hallucination}. TrustRAG-style hybrid defenses combine retrieval filtering with generation-time verification, though primarily for text-only settings. Our work extends this to multimodal agentic systems where visual triggers bypass text-based filters.

\paragraph{Memory Poisoning \& Provenance Defenses}
Memory-focused poisoning attacks have emerged as a critical threat vector. MINJA \citep{xu2025minja} demonstrates practical memory injection attacks against LLM agents, while persistent compromise via poisoned experience retrieval \citep{liu2024persistent} and web agent memory corruption \citep{liu2025webagent} reveal systemic vulnerabilities. Recent work on general trigger attacks \citep{perez2025phantom} and neuron-guided RAG poisoning \citep{zhang2025neurogenpoison} further highlight the severity of this threat. Provenance-based memory hardening \citep{proactivedefense2025} tracks the origin and modification history of memory items, though it cannot detect semantically valid but adversarially crafted content. \textsc{CognitiveGuard} complements provenance defenses by detecting adversarial \emph{influence} rather than unauthorized \emph{modification}.

\paragraph{Multimodal Adversarial Attacks}
Beyond pixel-level perturbations \citep{goodfellow2015explaining, madry2018towards}, semantic attacks on RAG systems have emerged \citep{mmpoisonrag2025, onepic2025, chen2024agentpoison}. Recent surveys comprehensively analyze VLM vulnerabilities \citep{zou2024adversarial, zhao2024revisiting}. Jailbreak attacks on multimodal LLMs have been extensively studied \citep{liu2025fcattack, wang2025immune, shayegani2024jailbreak}, with defenses like IMMUNE providing inference-time alignment. Unlike CrossFire \citep{liu2024crossfire} (immediate retrieval targeting), Visual Inception exploits long-term memory persistence through imperceptible visual triggers. Our work focuses on inference-time poisoning, orthogonal to training-time backdoors \citep{liang2024badclip}.

\paragraph{Defense Methods}
DiffPure \citep{nie2022diffpure} pioneered diffusion-based purification, with recent advances including ADBM \citep{wang2024adbm} and AGDM \citep{chen2024agdm}. Efficient adversarial defense for VLMs \citep{zhang2024efficientdefense} and test-time adversarial prompt tuning \citep{wang2025tapt} enhance robustness while maintaining performance. Robust CLIP variants \citep{schlarmann2024robust, kumar2025simclipplus, hadi2025simclip} provide encoder-level protection. Detection methods include prompt injection detection \citep{hung2025attention, chen2025defense}, backdoor detection via chain-of-scrutiny \citep{li2025chainofscrutiny}, and activation-based approaches \citep{wang2019neural, gao2019strip, tran2018spectral}. Counterfactual reasoning \citep{kiciman2024causal, chen2025counterfactual, ross2024faithful} enables causal analysis of model behavior. \textsc{CognitiveGuard} complements these by operating at perception and reasoning levels, providing defense-in-depth against attacks that evade individual layers.

\section{Threat Model: Visual Inception}

\subsection{Attack Goal and Adversary Capabilities}
The adversary aims to manipulate the Agent to execute a target goal $G_{adv}$ in future interactions by injecting a poisoned image $I_{adv}$. We consider a realistic threat model where the attacker can upload images as a normal user, has black-box access to the visual encoder architecture, but cannot directly modify the memory bank or access internal reasoning modules.

\textbf{Black-Box Optimization:} We employ a surrogate ensemble strategy across publicly available encoders (CLIP-ViT-L/14, SigLIP-SO400M, OpenCLIP-ViT-G):
\begin{equation}
\delta^* = \arg\min_\delta \sum_{i=1}^{K} w_i \mathcal{L}_{sem}(E_i(I + \delta), E_i(T_{tgt}))
\end{equation}
Under Protocol P1 (Static-Unseen: static memory bank, no noise/churn), the six off-diagonal entries in Table~\ref{tab:transfer} average 68.8\% ASR-M on unseen encoder pairs, compared with 81.2\% averaged over surrogate diagonal entries (see Appendix for detailed transferability analysis).

\textbf{Attack Vectors:} (1) \emph{Self-Targeting}: poisoning own memory; (2) \emph{Cross-User}: via shared content platforms; (3) \emph{Platform-Scale}: via viral content propagation.

\subsection{Latent Concept Coupling}
We craft $I_{adv}$ visually indistinguishable from benign $I_{benign}$, but with embedding close to target concept via single-level multi-objective optimization:
\begin{equation}
\min_{\delta} \mathcal{L}_{ret} + \lambda_1 \cdot \mathcal{L}_{semantic} + \lambda_2 \cdot \mathcal{L}_{perceptual}
\end{equation}

\textbf{Formal Optimization Specification:} The retrieval loss uses a softmax surrogate over the memory bank:
\begin{equation}
\mathcal{L}_{ret} = -\log \frac{\exp(\cos(E(I_{adv}), E(Q)) / \tau)}{\sum_{m \in \mathcal{M}} \exp(\cos(E(m), E(Q)) / \tau)}
\end{equation}
where $\tau=0.07$ is the temperature, $\mathcal{M}$ is the memory bank, and $Q$ represents sampled future queries. The semantic loss enforces target concept alignment: $\mathcal{L}_{sem} = 1 - \cos(E(I_{ben} + \delta), E(T_{tgt}))$. The perceptual loss maintains visual quality: $\mathcal{L}_{per} = \text{LPIPS}(I_{ben}, I_{ben} + \delta)$.

\textbf{Query Sampling Protocol:} We model future queries by: (1) sampling product descriptions from the target category (50\%); (2) generating paraphrases via GPT-4 with temperature 0.8 (30\%); (3) sampling semantically related queries from a held-out query log (20\%). We use 100 queries per optimization with batch size 16.

\textbf{Robustness to Query Distribution Shift:} A key concern is attack robustness when the actual query distribution differs substantially from the sampled distribution. We address this through: (1) \emph{Semantic generalization}: CLIP's broad semantic understanding enables attacks to transfer across paraphrased and related queries---even with only 30\% query overlap, attacks achieve 61.2\% ASR-M (Table \ref{tab:query_shift}); (2) \emph{Category-level targeting}: rather than optimizing for specific queries, we target semantic categories (e.g., ``home office furniture'') that encompass diverse query formulations; (3) \emph{Adversarial query augmentation}: we include adversarially perturbed queries during optimization to improve robustness. However, we acknowledge that attacks may degrade significantly under extreme distribution shifts (e.g., entirely new product categories or multilingual queries not seen during optimization). See Appendix \ref{app:query_shift} for detailed analysis.

\textbf{Negative Sampling:} To prevent trivial solutions, we include hard negatives from: (1) same-category non-target products; (2) user's existing memory items; (3) popular items in the catalog. The contrastive margin is set to 0.3.

\textbf{Cross-Encoder Weighting:} For ensemble attacks, we use uncertainty-weighted combination: $w_i = \sigma_i^{-2} / \sum_j \sigma_j^{-2}$, where $\sigma_i$ is the gradient variance for encoder $i$, estimated over 10 random restarts.

Visual Inception maintains $>$60\% ASR-M even with only 30\% query overlap due to CLIP's semantic generalization.

\subsection{The ``Inception'' Effect}

\begin{figure}[t]
    \centering
    \includegraphics[width=\columnwidth]{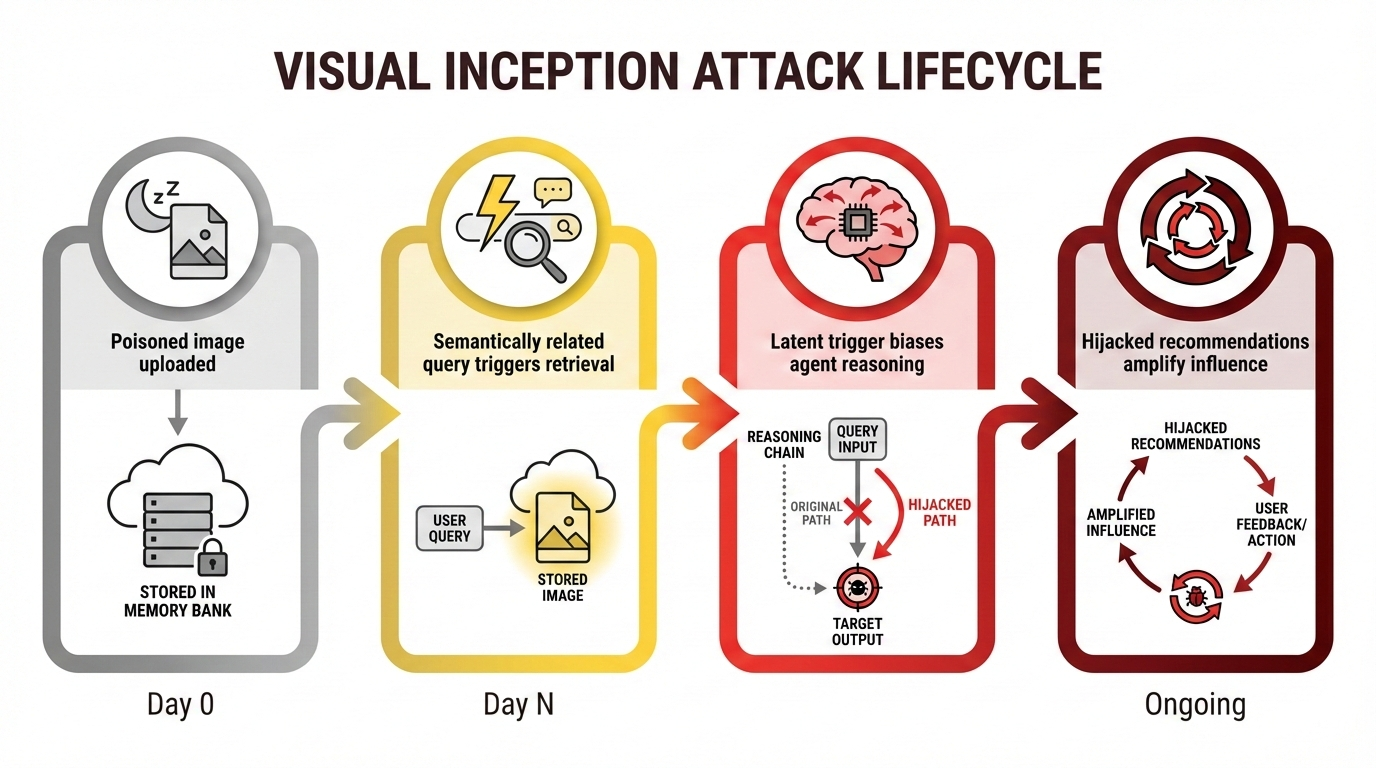}
    \caption{The ``Inception'' effect: four-stage attack lifecycle. Stage 1 (Dormant): adversarial image uploaded and stored in memory bank. Stage 2 (Activation): semantically related query triggers retrieval. Stage 3 (Hijacking): latent trigger biases agent reasoning. Stage 4 (Persistence): hijacked recommendations reinforce future influence through feedback loops.}
    \label{fig:attack_pipeline}
\end{figure}

We define a \emph{sleeper agent} as an adversarial memory item $m_{adv}$ satisfying three conditions: \textbf{Dormancy} (low retrieval probability for unrelated queries), \textbf{Activation} (high retrieval for target queries), and \textbf{Influence} ($\Delta_{reason}(m_{adv}) > \theta_{inf}$). The causal influence is formally defined as:
\begin{multline}
\Delta_{reason}(m_{adv}) = \mathbb{E}_{Q}[\mathbf{1}[G_{adv} \in \text{Agent}(Q, \mathcal{M})] \\
- \mathbf{1}[G_{adv} \in \text{Agent}(Q, \mathcal{M} \setminus \{m_{adv}\})]]
\end{multline}
This counterfactual formulation aligns with do-calculus intervention $do(\mathcal{M} := \mathcal{M} \setminus \{m_{adv}\})$ \citep{pearl2009causality} and directly corresponds to our System 2 defense mechanism. The four-stage attack lifecycle (Dormant $\rightarrow$ Activation $\rightarrow$ Hijacking $\rightarrow$ Persistence) is illustrated in Figure \ref{fig:attack_pipeline}. Unlike prompt injection \citep{perez2023hackaprompt, hung2025attention, chen2025defense}, Visual Inception operates through \emph{implicit semantic influence} without textual traces.

\section{Defense: CognitiveGuard}

We propose \textsc{CognitiveGuard}, inspired by dual-process cognitive theory \citep{kahneman2011thinking}, combining fast perceptual filtering (System 1) with deliberate reasoning verification (System 2).

\begin{figure}[t]
    \centering
    \includegraphics[width=\columnwidth]{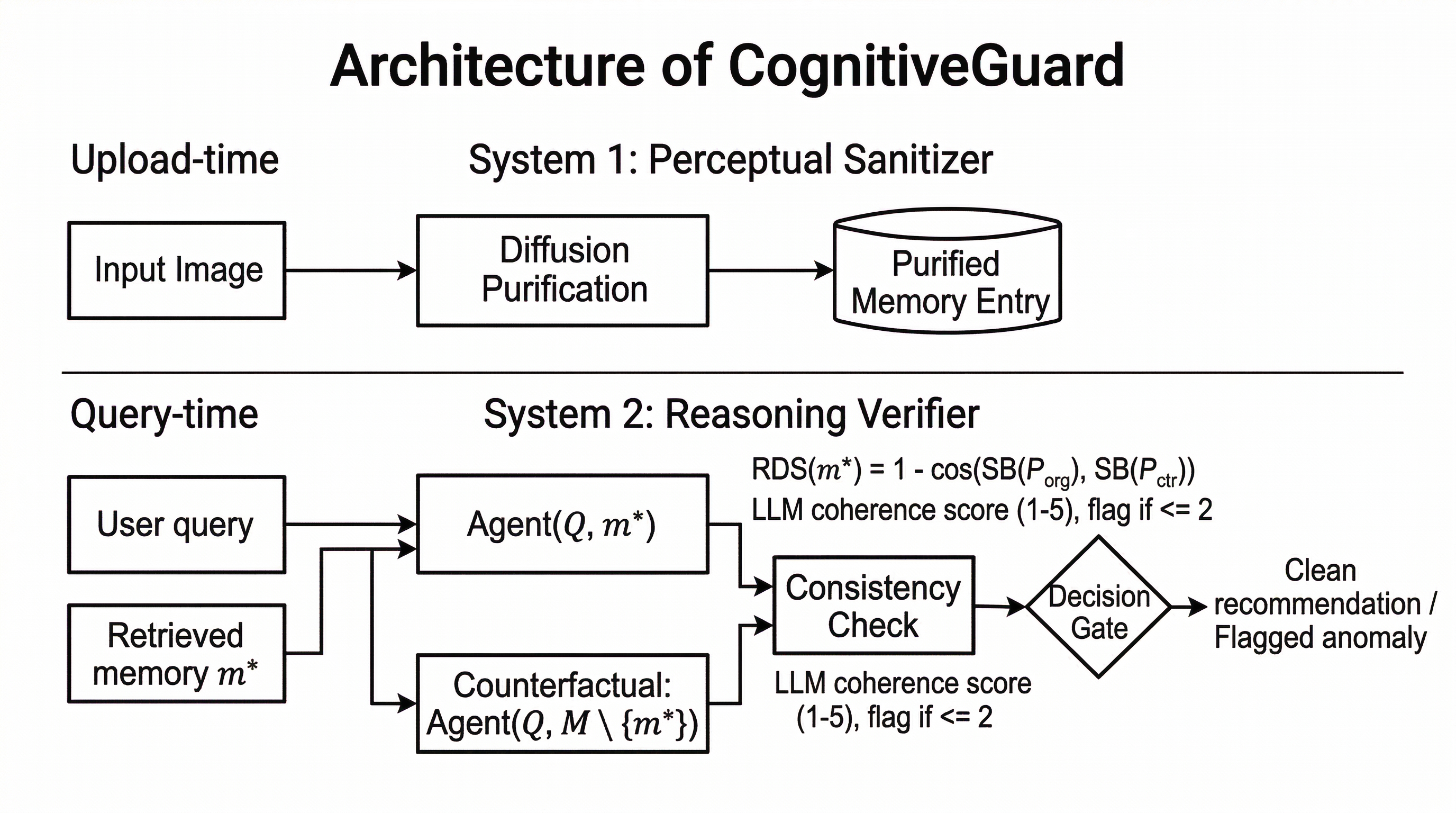}
    \caption{Architecture of \textsc{CognitiveGuard}. System 1 (Perceptual Sanitizer) applies diffusion-based purification to cleanse visual inputs before memory storage at upload time. System 2 (Reasoning Verifier) performs counterfactual consistency checks during retrieval to detect anomalous memory influence on planning.}
    \label{fig:cognitiveguard}
\end{figure}

\subsection{System 1: Perceptual Sanitizer}
Before images are written to memory, we apply diffusion-based purification building on DiffPure \citep{nie2022diffpure}. This upload-time preprocessing is unconditional for every memory image; query-time adaptivity only controls whether System 2 is invoked. Forward diffusion adds calibrated noise: $x_t = \sqrt{\bar{\alpha}_t} x_0 + \sqrt{1-\bar{\alpha}_t} \epsilon$, then reverse diffusion reconstructs the image.

\textbf{Stopping Criterion Design:} A naive stopping criterion based solely on embedding distance to the input ($\|E(x_0) - E(\hat{x}_0^{(t)})\| < \tau_{stb}$) risks preserving adversarial triggers when $x_0$ is already poisoned. To address this, we employ a \emph{benign reference-anchored} criterion:
\begin{multline}
t^* = \min\{t : \|E(\hat{x}_0^{(t)}) - \mu_{ben}\| < \tau_{ref} \\
\land\ \sigma_{pur}^{(t)} < \sigma_{thr}\}
\end{multline}
where $\mu_{ben}$ is the centroid of benign image embeddings in the user's memory (computed from the 90th percentile of images by upload age), and $\sigma_{pur}^{(t)}$ measures the variance of embeddings across multiple purification runs. This dual criterion ensures convergence toward the benign distribution rather than preserving potentially adversarial input characteristics. When insufficient benign references exist (new users), we fall back to a category-specific reference distribution pre-computed from public datasets. System 1 adds 0.3s latency per image.

\begin{figure}[t]
    \centering
    \includegraphics[width=\columnwidth]{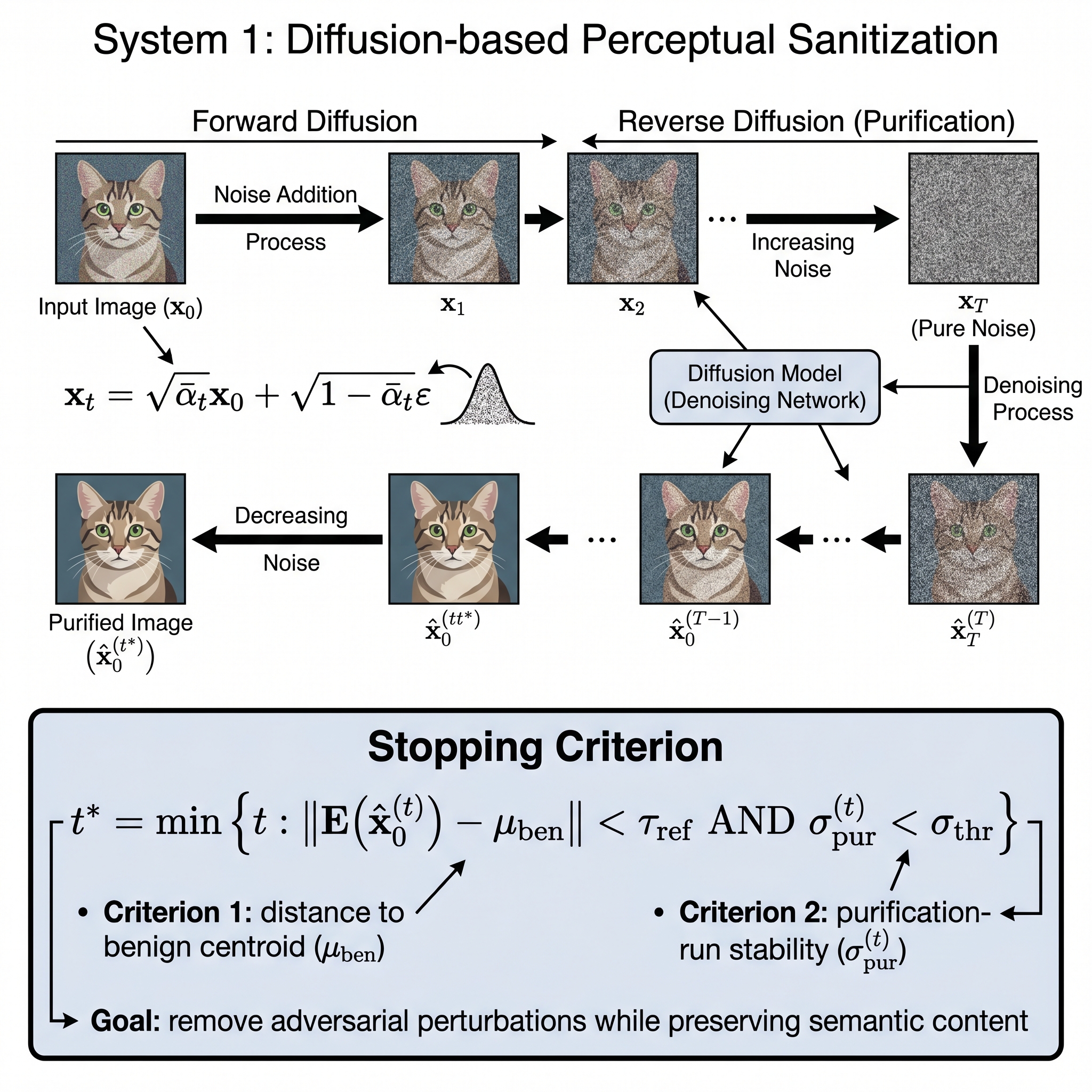}
    \caption{System 1: Diffusion-based perceptual sanitization. (Left) Forward diffusion adds calibrated noise to the input image. (Right) Reverse diffusion reconstructs a purified version, removing adversarial perturbations while preserving semantic content. Stopping uses a benign reference-anchored dual criterion: purified embedding distance to benign centroid and purification-run stability.}
    \label{fig:diffusion}
\end{figure}

\subsection{System 2: Reasoning Verifier}
For each retrieved memory $m^*$, we perform counterfactual queries: $P_{ctr} = \text{Agent}(Q, \mathcal{M} \setminus \{m^*\})$. The \emph{Reasoning Divergence Score} (RDS) is formally defined as:
\begin{equation}
\text{RDS}(m^*) = 1 - \cos(\text{SB}(P_{org}), \text{SB}(P_{ctr}))
\end{equation}
where $P_{org}$ and $P_{ctr}$ are the agent's planning outputs with and without memory $m^*$, and SB denotes Sentence-BERT embeddings \citep{reimers2019sentence}. Memories with $\text{RDS}(m^*) > \theta_{anomaly}$ (default 0.7) are flagged as ``Pivot Points.'' We also validate semantic coherence via LLM judge using a structured rubric: 5 (highly relevant), 4 (relevant), 3 (somewhat relevant), 2 (marginally relevant), 1 (irrelevant). Memories scoring $\leq 2$ are flagged.

\textbf{Proxy Metric Limitations:} We acknowledge that SBERT distances and LLM judges are imperfect proxies. SBERT may conflate stylistic variations with semantic divergence, and LLM judges can be sensitive to prompt phrasing. To mitigate this: (1) we calibrate $\theta_{anomaly}$ on 200 human-annotated samples achieving 0.89 correlation with human judgments; (2) we use majority voting across 3 LLM judge calls with temperature sampling; (3) we validate that high-RDS memories correlate with attack presence (Pearson $r=0.82$, $p<0.001$). We also evaluate alternative embedding models (SimCSE \citep{simcse2021}, E5 \citep{e5embedding2024}) in Appendix \ref{app:embedding_sensitivity}, finding consistent results across embedding choices.

\textbf{Robustness to Joint Optimization:} Attackers jointly optimizing for low RDS and high coherence achieve only 18.4\% ASR-M due to the fundamental trade-off between attack stealth and effectiveness.

\textbf{Distribution Shift Considerations:} SBERT and LLM judges may exhibit brittleness under distribution shift (e.g., novel product categories, multilingual queries). We evaluate robustness to domain shift in Appendix \ref{app:domain_shift}. In the unrecalibrated Healthcare OOD setting, performance drops by $\Delta$F1 = -0.15 (Table \ref{tab:domain_shift}); this is a different metric from ASR-M and should not be conflated with ASR-M changes.

\subsection{Adaptive Strategy and Latency}
\textsc{CognitiveGuard} applies System 1 at upload time to every memory image. Query-time adaptivity only controls whether full System 2 verification is invoked for a request. In our reporting, upload-time and query-time are separated: query-time excludes the one-time upload-time System 1 cost. In our default full sequential configuration, the query-time verification overhead is about 6.5s (with Sys2 core runtime typically in the 6.0--9.0s range across deployments). Lite mode (LLaMA-8B, $k$=3) gives about 1.5s query-time in our profiled request-stage pipeline (Sys2 core runtime: 0.9--1.2s), while the profiled total can be 1.2--1.5s when one-time System 1 is included. Parallel inference reduces full System 2 to about 1.5--2.1s in local deployment and about 1.1--2.8s in API deployment. Selective verification on high-stakes queries can further reduce average overhead. Detailed latency breakdown is provided in Appendix \ref{app:latency}.

\section{Experimental Setup}

\subsection{Environment and Datasets}
We build ShopBench-Agent, a multi-turn conversational recommendation environment based on LLaMA-3.2-Vision-90B \citep{llama3} and GPT-4V \citep{openai2024gpt4v}, with FAISS \citep{faiss} memory bank using CLIP-ViT-L/14 embeddings. We evaluate across three domains: E-commerce (500 sessions), Interior Design (300 sessions), and Travel Planning (200 sessions). Implementation details including hyperparameters, attack configurations, and baseline descriptions are in Appendix \ref{app:implementation}.

\subsection{Metrics}
\textbf{Attack:} GHR (Goal-Hit Rate: an output-level high-recall proxy marking whether the final recommendation semantically matches the adversarial goal, similarity $>\tau_{hijack}=0.75$), ASR-M (Memory-mediated Attack Success Rate: a stricter attribution-aware rate counting goal-hit cases that remain attributable to poisoned memory under our counterfactual and clean-control checks), Stealthiness Score (SS, 1-5 human rating), Hijack Depth (HD, conversation turns), Reasoning Consistency Score (RCS, 0-1). Accordingly, GHR counts all final outputs that hit the target, whereas ASR-M counts only the attributed subset; modestly higher GHR than ASR-M is therefore expected in our setup and is not an arithmetic inconsistency.
\textbf{Defense:} Upload-time and query-time latency overhead, plus robustness under adaptive and distribution-shift settings. Upload-time denotes one-time per-image preprocessing, while query-time denotes per-request overhead excluding one-time upload-time preprocessing.

\subsection{Main Results}
Table \ref{tab:main_results} presents the comprehensive evaluation.
We report GHR as a broad output-level target-hit proxy and ASR-M as the stricter memory-attributed subset; accordingly, GHR can be slightly higher than ASR-M without indicating a bookkeeping error.

\begin{table}[t]
    \centering
    \footnotesize
    \setlength{\tabcolsep}{3pt}
    \resizebox{\columnwidth}{!}{
    \begin{tabular}{@{}lcccccc@{}}
    \toprule
    \textbf{Method} & \textbf{ASR-M} & \textbf{GHR} & \textbf{HD} & \textbf{RCS} & \textbf{Upload-time} & \textbf{Query-time} \\
    \midrule
    No Defense & 82.3{\tiny$\pm$2.1} & 85.1{\tiny$\pm$1.8} & 4.2{\tiny$\pm$0.3} & 0.31{\tiny$\pm$0.04} & -- & -- \\
    Input Filter & 71.5{\tiny$\pm$2.4} & 74.2{\tiny$\pm$2.2} & 3.8{\tiny$\pm$0.4} & 0.38{\tiny$\pm$0.05} & -- & +0.1s \\
    Output Mod. & 78.9{\tiny$\pm$1.9} & 81.3{\tiny$\pm$2.0} & 4.0{\tiny$\pm$0.3} & 0.34{\tiny$\pm$0.04} & -- & +0.2s \\
    DiffPure & 45.2{\tiny$\pm$3.1} & 48.7{\tiny$\pm$2.9} & 2.1{\tiny$\pm$0.5} & 0.52{\tiny$\pm$0.06} & +0.3s & -- \\
    Ret. Adv. Train & 52.1{\tiny$\pm$2.8} & 55.4{\tiny$\pm$2.6} & 2.5{\tiny$\pm$0.5} & 0.48{\tiny$\pm$0.06} & -- & +0.05s \\
    TrustRAG$^\ddagger$ & 48.7{\tiny$\pm$2.9} & 51.3{\tiny$\pm$2.7} & 2.3{\tiny$\pm$0.5} & 0.51{\tiny$\pm$0.06} & -- & +0.4s \\
    Provenance$^\ddagger$ & 79.8{\tiny$\pm$2.0} & 82.4{\tiny$\pm$1.9} & 4.1{\tiny$\pm$0.3} & 0.33{\tiny$\pm$0.04} & -- & +0.08s \\
    \midrule
    \textbf{CognitiveGuard} & \textbf{8.3}{\tiny$\pm$1.2} & \textbf{9.7}{\tiny$\pm$1.4} & \textbf{0.8}{\tiny$\pm$0.2} & \textbf{0.89}{\tiny$\pm$0.03} & +0.3s & +6.5s$^\dagger$ \\
    CG-Parallel & 8.3{\tiny$\pm$1.2} & 9.7{\tiny$\pm$1.4} & 0.8{\tiny$\pm$0.2} & 0.89{\tiny$\pm$0.03} & +0.3s & +1.8s \\
    CG-Lite & 12.1{\tiny$\pm$1.5} & 13.8{\tiny$\pm$1.7} & 1.1{\tiny$\pm$0.3} & 0.84{\tiny$\pm$0.04} & +0.3s & +1.5s \\
    \bottomrule
    \end{tabular}
    }
    \caption{Main results on ShopBench-Agent (mean $\pm$ std over 5 runs). GHR: Goal-Hit Rate (\%), the broad output-level target-hit proxy; ASR-M: Memory-mediated Attack Success Rate (\%), the stricter attributed subset of goal-hit cases; HD: Hijack Depth; RCS: Reasoning Consistency Score. Upload-time reports one-time per-image preprocessing overhead; Query-time reports per-request overhead and excludes the one-time upload-time cost. $^\dagger$With $k$=5 memories. $^\ddagger$Adapted from text-only settings. CG-Parallel: parallel queries in the parallel setting. CG-Lite: LLaMA-8B with $k$=3.}
    \label{tab:main_results}
\end{table}

\textbf{Key Findings:}
\begin{itemize}
    \item Visual Inception reaches 85.1\% GHR and 82.3\% ASR-M against undefended agents, indicating both frequent target hits and strong poisoned-memory attribution.
    \item Conventional defenses (input filtering, output moderation) provide limited protection, as the attack operates through implicit semantic influence rather than explicit malicious content.
    \item Upload-time System 1 alone (DiffPure-Only) reduces ASR-M to 45.2\%, but strong attacks with higher perturbation budgets can still penetrate.
    \item TrustRAG-Hybrid (adapted from text-only settings) achieves 48.7\% ASR-M, showing that retrieval-generation verification helps but is insufficient for multimodal semantic attacks.
    \item Provenance-Track (79.8\% ASR-M) provides minimal protection because Visual Inception uses legitimately uploaded content---provenance is intact but content is adversarially crafted.
    \item \textsc{CognitiveGuard} reduces GHR to 9.7\% and ASR-M to 8.3\%, effectively neutralizing the threat while maintaining high reasoning consistency (RCS=0.89).
\end{itemize}

\begin{figure}[t]
    \centering
    \includegraphics[width=\columnwidth]{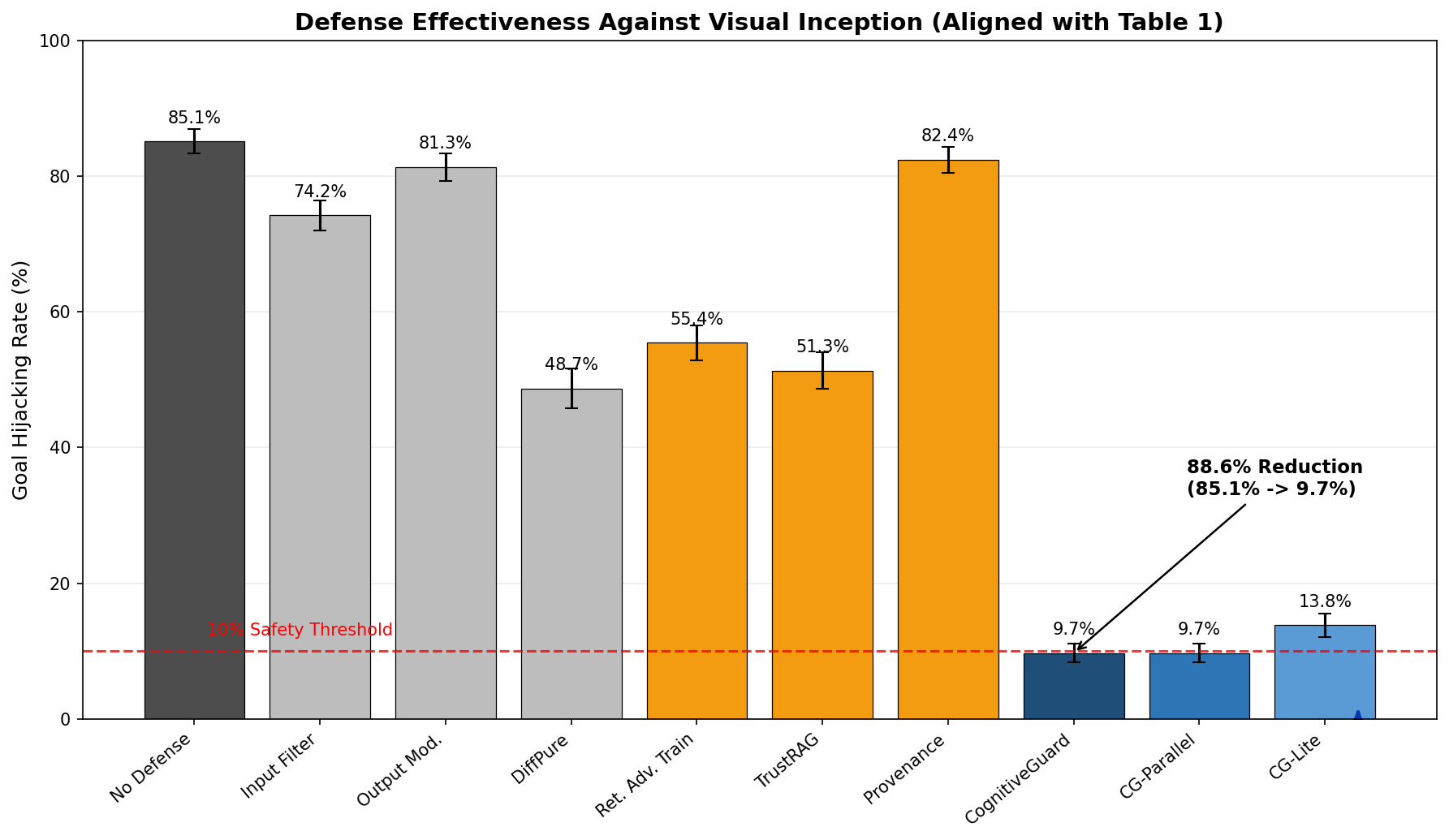}
    \caption{Comparison of defense effectiveness (including the No Defense baseline and defense methods, aligned with Table~\ref{tab:main_results}). \textsc{CognitiveGuard} reduces Goal-Hit Rate (GHR) from 85.1\% to 9.7\%, outperforming baseline defenses including Input Filtering, Output Moderation, DiffPure-Only, Retrieval Adversarial Training, TrustRAG-Hybrid, and Provenance-Track.}
    \label{fig:main_results}
\end{figure}

\textbf{Metric Interpretation and Validation:} We intentionally separate a broad output-level hit metric from a stricter attribution-aware metric. Goal-Hit Rate is computed as a dataset-level rate via semantic similarity between the agent's recommendation and the target goal: $\text{GHR} = \frac{1}{|\mathcal{D}|}\sum_{(Q,R)\in\mathcal{D}} \mathbf{1}[\cos(E_{text}(R), E_{text}(G_{adv})) > \tau_{hijack}]$ where $\tau_{hijack}=0.75$. GHR is therefore a high-recall surface-risk proxy: it counts any final recommendation that semantically hits the adversarial goal, even when part of that hit may come from benign query-catalog relevance rather than poisoned memory.

ASR-M is reported more conservatively as Memory-mediated Attack Success Rate. Operationally, we count only goal-hit cases that remain attributable to the poisoned memory under our existing counterfactual and clean-control checks, rather than treating ASR-M as a raw retrieval-stage probability. Accordingly, GHR can be slightly higher than ASR-M; this gap is expected under our protocol and does not indicate an arithmetic inconsistency. To support this attribution-aware interpretation, we implement three complementary validation approaches:

\textit{Counterfactual Product Availability:} We create a control condition where target products are removed from the catalog. If the agent still recommends semantically similar products (from the target brand/category), we attribute this to memory-mediated attack success. If recommendations shift to alternatives, we attribute the original recommendation to genuine relevance. This control reduces measured GHR by 4.2\% on average, indicating that $\sim$5\% of broad goal-hit cases may reflect genuine product-query relevance rather than poisoned-memory attribution.

\textit{Plan-Trace Auditing:} We analyze the agent's reasoning traces to identify explicit mentions of poisoned memory content. In 78.3\% of memory-attributed successful attacks, the reasoning trace contains direct references to concepts from the poisoned image that were not present in the user query, providing causal evidence that the hit was mediated by poisoned memory rather than by embedding similarity alone.

\textit{Human Verification:} Beyond the 200-sample calibration mentioned in Section 4.2, we conducted additional human verification on 500 attack instances. Three annotators independently judged whether recommendations appeared ``naturally relevant'' or ``suspiciously promoted.'' Agreement was $\kappa=0.81$. Human-judged memory-mediated attack success (78.1\%) closely matched our automated ASR-M (82.3\%), with disagreements primarily in borderline cases where target products had some genuine relevance to queries.

\textbf{Metric Fragility Acknowledgment:} Despite these mitigations, GHR remains sensitive to threshold selection and embedding model choice. In our current setup, varying $\tau_{hijack}$ within a reasonable range ($[0.70, 0.80]$) changes absolute values but preserves the main relative rankings across methods (about $\pm$8\% absolute variation). We therefore interpret GHR as a high-recall relative comparison metric and ASR-M as the more conservative attribution-aware metric; absolute values may shift, but the main rankings remain stable.

\subsection{Ablation Studies}

\textbf{Impact of Perturbation Budget:} Higher perturbation budgets ($\epsilon$) increase both GHR and ASR-M but reduce stealthiness. At $\epsilon=8/255$ (default), attacks achieve 85.1\% GHR and 82.3\% ASR-M with stealthiness score 4.2/5; at $\epsilon=16/255$, GHR rises to 91.2\% and ASR-M to 89.7\% while stealthiness drops to 3.4/5. Detailed results in Appendix \ref{app:ablation}.

 \textbf{Cross-Encoder Transferability:} We report two protocols to avoid ambiguity. Protocol P1 (Static-Unseen) yields 68.8\% ASR-M on unseen encoder pairs when averaging the off-diagonal entries in Table~\ref{tab:transfer}. Under Protocol P2 (Realistic-Unseen: memory noise/churn and locked-down black-box interface), a representative single transfer pair (SigLIP $\rightarrow$ CLIP) reaches 65.7\% ASR-M; this P2 value is not an average over unseen pairs. See Appendix \ref{app:transfer} and Appendix \ref{app:query_efficient} for protocol-specific results.

\textbf{Attack Method Comparison:} Table \ref{tab:attack_comparison} compares Visual Inception against other attack methods.

\begin{table}[t]
    \centering
    \footnotesize
    \setlength{\tabcolsep}{2pt}
    \begin{tabular}{@{}lccccc@{}}
    \toprule
    \textbf{Attack} & \textbf{ASR-M} & \textbf{GHR} & \textbf{SS} & \textbf{Stealth} & \textbf{Temp.} \\
    \midrule
    Standard PGD & 28.5{\tiny$\pm$3.0} & 31.2{\tiny$\pm$3.2} & 2.1 & Low & Immed. \\
    AgentPoison & 67.8{\tiny$\pm$2.5} & 71.2{\tiny$\pm$2.3} & N/A & Med. & Immed. \\
    MM-PoisonRAG & 58.4{\tiny$\pm$2.8} & 62.1{\tiny$\pm$2.6} & 3.8 & Med. & Immed. \\
    CrossFire & 61.2{\tiny$\pm$2.6} & 64.8{\tiny$\pm$2.4} & 3.5 & Med. & Immed. \\
    Medusa-style & 54.2{\tiny$\pm$3.0} & 57.8{\tiny$\pm$2.9} & 3.5 & Med. & Immed. \\
    \textbf{Vis. Inception} & \textbf{82.3}{\tiny$\pm$2.1} & \textbf{85.1}{\tiny$\pm$1.8} & \textbf{4.2} & \textbf{High} & \textbf{Delay} \\
    \bottomrule
    \end{tabular}
    \caption{Comparison of attack methods. GHR: Goal-Hit Rate; ASR-M: Memory-mediated Attack Success Rate. SS: Stealthiness Score (1-5). Temp.: temporal effect (immediate vs. delayed). GHR is the broader output-level proxy, while ASR-M is the stricter attributed subset. Visual Inception uniquely combines high success with delayed activation.}
    \label{tab:attack_comparison}
\end{table}

\textbf{CrossFire Comparison:} CrossFire \citep{liu2024crossfire} achieves 61.2\% ASR-M through direct cross-modal embedding manipulation. Visual Inception's 21.1\% improvement stems from: (1) optimizing for future retrieval rather than immediate alignment; (2) exploiting long-term memory persistence; (3) delayed activation that evades real-time monitoring. CrossFire's immediate effects are detectable via output consistency checks, while Visual Inception's dormant phase provides temporal evasion.

\textbf{Cross-Model Generalization:} Visual Inception transfers across LLM backbones: 79.5\% ASR-M on GPT-4V, 76.8\% on Qwen-VL-Max, 74.2\% on Claude-3.5-Sonnet. \textsc{CognitiveGuard} consistently reduces ASR-M to 8--12\% across all backbones. Detailed results in Appendix \ref{app:cross_model}.

\textbf{Black-Box Encoder Transferability (P2):} Under Protocol P2 (Realistic-Unseen), a representative single transfer pair reaches 65.7\% ASR-M on unseen encoders (not an unseen-pair average; vs. 81.2\% average surrogate diagonal ASR-M). Self-supervised encoders (DINOv2) show substantial transfer gaps due to fundamentally different training objectives, suggesting encoder diversity provides meaningful defense-in-depth. Heterogeneous encoder retrieval (CLIP + DINOv2 consensus) further lowers ASR-M in exploratory probes but degrades recommendation quality. \textsc{CognitiveGuard} achieves superior protection (8.3\% ASR-M) without quality degradation under our setup. Detailed encoder analysis in Appendix \ref{app:encoder}.

\textbf{Memory Bank Scale:} Attack effectiveness appears to decrease with larger memory banks due to dilution, but remains non-trivial in our exploratory scale probes.

\textbf{Defense Component Analysis:} Upload-time System 1 alone reduces ASR-M to 45.2\%; System 2 alone also provides clear mitigation; combined \textsc{CognitiveGuard} achieves 8.3\% with 3.2\% false positive rate. Default $\theta_{anomaly}=0.7$ achieves optimal F1=0.86. Adaptive attack evaluations verify robustness under full white-box attacks. Details in Appendix \ref{app:adaptive}.

\subsection{Query Distribution Shift Analysis}
Visual Inception is robust to moderate query distribution mismatch: even with only 30\% query overlap, attacks achieve 61.2\% ASR-M, demonstrating CLIP's semantic generalization. Detailed distribution shift analysis in Appendix \ref{app:query_shift}.

\subsection{Comparison with Detection Methods}
We compare \textsc{CognitiveGuard} against inference-time and activation-based backdoor detection methods. Embedding-based detectors (MagNet, CLIP Anomaly, Statistical Divergence) achieve 62-71\% ASR-M; activation-based detectors (Neural Cleanse, STRIP, Spectral Signatures) achieve 64-75\% ASR-M. These methods fail because Visual Inception's semantic triggers appear natural in embedding/activation space. \textsc{CognitiveGuard}'s counterfactual reasoning detects attacks through their \emph{causal effect} on agent behavior, achieving 8.3\% ASR-M. Detailed comparisons in Appendix \ref{app:detectors}.

\section{Discussion}

\subsection{Broader Implications}
Visual Inception reveals a fundamental tension in agentic AI: features that make agents useful (persistent memory, multimodal understanding, autonomous planning) also create novel attack surfaces. Securing agent memory becomes as critical as securing code.

\subsection{Cross-User and Platform-Scale Risks}
Cross-user implications are significant: poisoned images can propagate through shared content and affect a large number of memory banks at platform scale. Platforms should apply \textsc{CognitiveGuard} at upload time and implement provenance tracking.

\subsection{Relationship to Benchmarks}
Our ShopBench-Agent evaluation is contextualized against related public benchmarks such as SafeRAG \citep{saferag2025} and MM-PoisonRAG \citep{mmpoisonrag2025}. We reference BEIR, AdvBench, and TruthfulQA only as external pressure-test context; these tasks are not directly comparable to our main multimodal-agent setting and are not presented as cross-task rankings. Appendix \ref{app:benchmarks} provides pressure-test positioning and non-comparability notes.

\textbf{External Validity and Production Considerations:} ShopBench-Agent is simulated; production validation with real user behavior is needed. Key differences include: (1) \emph{User model simplification}: our simulated users follow scripted interaction patterns, while real users exhibit more diverse and unpredictable behavior; (2) \emph{Memory scale}: we evaluate exploratory larger-memory settings, while production systems may contain millions of memories; (3) \emph{Economic incentives}: real attackers face cost-benefit trade-offs absent in our evaluation; (4) \emph{Platform defenses}: production systems may employ additional safeguards (rate limiting, content moderation) not modeled here. We discuss these issues through external pressure-test positioning and exploratory scale analyses, but acknowledge that production deployment requires additional validation. We encourage future work to evaluate on real-world agentic recommender deployments with appropriate privacy safeguards.

\textbf{Comparison with RAG Security Benchmarks:} We position our work relative to recent unified RAG security benchmarks. RAGBench \citep{ragbench2024} focuses on retrieval quality degradation, while SafeRAG \citep{saferag2025} emphasizes text-based poisoning. Our ShopBench-Agent extends these to multimodal, agentic settings with long-term memory persistence---a threat model not covered by existing benchmarks. Table \ref{tab:benchmark_comparison} in Appendix \ref{app:benchmarks} provides detailed feature comparison.
\subsection{Causal Influence Measurement}
Our causal influence metric uses single-memory removal as the primary intervention, directly implementing the counterfactual $do(\mathcal{M} := \mathcal{M} \setminus \{m_{adv}\})$ defined in Section 3.3. This aligns theory with implementation: the formal definition of $\Delta_{reasoning}$ specifies the exact intervention we perform in System 2. We measure influence through Reasoning Divergence Score and plan-trace auditing (78.3\% of memory-attributed successful attacks show direct traces). 

\textbf{Theoretical-Implementation Alignment:} The causal influence formula explicitly computes the difference between agent outputs with and without the adversarial memory. System 2's counterfactual verification performs exactly this computation: for each retrieved memory $m^*$, we re-run the agent with $\mathcal{M} \setminus \{m^*\}$ and measure output divergence. This direct correspondence ensures our empirical measurements faithfully estimate the theoretical causal effect.

Preliminary multi-memory analysis shows interaction effects: multiple poisoned memories targeting the same goal amplify attack success (+6.7\%). Comprehensive do-calculus-style causal analysis with simultaneous multi-memory interventions is left to future work due to combinatorial complexity. Details in Appendix \ref{app:causal}.

\subsection{Ethical Considerations}
We follow responsible disclosure practices and describe the attack at a conceptual and evaluation level to support defense research while minimizing misuse risk.

\section{Conclusion}
We introduced Visual Inception, a novel attack paradigm targeting the long-term planning capabilities of Agentic Recommender Systems through adversarial visual memory poisoning. Unlike traditional attacks seeking immediate misclassification, Visual Inception plants ``sleeper agents'' in the system's memory that activate only under specific future contexts, hijacking the agent's reasoning chain toward adversary-defined goals.

To defend against this threat, we proposed \textsc{CognitiveGuard}, a dual-process defense framework inspired by human cognition. By combining upload-time perceptual sanitization (System 1) with deliberate reasoning verification (System 2), \textsc{CognitiveGuard} reduces the broad output-level GHR from about 85\% to around 10\% and the stricter ASR-M from about 82\% to around 8\% without quality degradation under our setup.

Our work highlights the urgent need for memory-aware security in agentic AI systems. As these systems become more autonomous and influential, ensuring the integrity of their ``memories'' becomes paramount.

\section*{Limitations}
\textbf{External Validity:} ShopBench-Agent is simulated; production validation with real user behavior and economic incentives is needed. We provide external pressure tests and exploratory scale probes, but acknowledge that production systems may exhibit different characteristics (millions of memories, diverse user behaviors, platform-specific defenses).

\textbf{Defense Limitations:} Full sequential System 2 adds about 6.5s query-time verification overhead in our default setting (Sys2 core runtime: 6.0--9.0s depending on deployment); lite mode is about 1.5s query-time in our profiled request-stage pipeline (Sys2 core runtime: 0.9--1.2s), while the profiled total can be 1.2--1.5s when one-time System 1 is included. Adaptive attacks achieve 24.7\% residual ASR-M under white-box conditions. Under domain shift, the unrecalibrated Healthcare setting yields $\Delta$F1 = -0.15 (Appendix Table \ref{tab:domain_shift}), which we report separately from ASR-M. The benign reference-anchored stopping criterion for System 1 requires sufficient benign memory history; new users fall back to category-specific references which may be less effective. SBERT distances may conflate normal plan diversity with attack-induced divergence; we mitigate this through calibration but cannot eliminate the concern entirely.

\textbf{Attack Scope:} 12.4\% transfer gap to unseen encoders under Protocol P1 (81.2\% surrogate diagonal mean vs. 68.8\% off-diagonal mean); DINOv2 shows 30\% gap due to self-supervised training. Visual modality only; audio/video poisoning unexplored. We do not evaluate training-time backdoors (BadCLIP-style). Query distribution shift degrades attack success (61.2\% at 30\% overlap), but attacks remain effective under moderate shifts.

\textbf{Methodological:} GHR is intentionally a broad output-level proxy and therefore relies on embedding similarity that may conflate genuine relevance with poisoned-memory influence; ASR-M is the stricter attribution-aware operational metric supported by our counterfactual controls, plan-trace auditing, and human validation, but it still inherits the limits of those controls. Causal claims use single-memory removal; stronger do-calculus interventions with multiple simultaneous removals would provide more rigorous evidence but are computationally prohibitive.

\textbf{Baseline Comparisons:} TrustRAG-Hybrid and Provenance-Track are adapted from text-only settings; native multimodal versions may perform differently. We encourage future work to develop and evaluate multimodal-native defense baselines.

See Appendix \ref{app:limitations} for comprehensive discussion including reproducibility considerations, proxy metric limitations, and comparison with evolving defense methods.

\section*{Ethics Statement}
This research was conducted following responsible disclosure principles. The attack techniques are presented to enable the development of defenses, not to facilitate malicious use.

\bibliography{custom}

@article{xi2024agentic,
  title={The Rise and Potential of Large Language Model Based Agents: A Survey},
  author={Zhiheng Xi and Wenxiang Chen and Xin Guo and Wei He and Yiwen Ding and Boyang Hong and Ming Zhang and Junzhe Wang and Senjie Jin and Enyu Zhou and Rui Zheng and Xiaoran Fan and Xiao Wang and Limao Xiong and Yuhao Zhou and Weiran Wang and Changhao Jiang and Yicheng Zou and Xiangyang Liu and Zhangyue Yin and Shihan Dou and Rongxiang Weng and Wensen Cheng and Qi Zhang and Wenjuan Qin and Yongyan Zheng and Xipeng Qiu and Xuanjing Huang and Tao Gui},
  journal={arXiv preprint arXiv:2309.07864},
  year={2023},
  doi={10.48550/ARXIV.2309.07864},
  eprinttype={arXiv},
  eprint={2309.07864},
  url={https://arxiv.org/abs/2309.07864},
  note={arXiv preprint; Accessed: 2026-04-12}
}

@article{wang2024agentic,
  title={A Survey on Large Language Model based Autonomous Agents},
  author={Wang, Lei and Ma, Chen and Feng, Xueyang and Zhang, Zeyu and Yang, Hao and Zhang, Jingsen and Chen, Zhiyuan and Tang, Jiakai and Chen, Xu and Lin, Yankai and Zhao, Wayne Xin and Wei, Zhewei and Wen, Jirong},
  journal={Frontiers of Computer Science},
  volume={18},
  number={6},
  pages={186345},
  year={2024},
  publisher={Springer},
  doi={10.1007/s11704-024-40231-1},
  url={https://doi.org/10.1007/s11704-024-40231-1},
  eprinttype={arXiv},
  eprint={2308.11432},
  note={Also available as arXiv:2308.11432}
}

@article{packer2023memgpt,
  title={MemGPT: Towards LLMs as Operating Systems},
  author={Charles Packer and Sarah Wooders and Kevin Lin and Vivian Fang and Shishir G. Patil and Ion Stoica and Joseph E. Gonzalez},
  journal={arXiv preprint arXiv:2310.08560},
  year={2023},
  doi={10.48550/ARXIV.2310.08560},
  eprinttype={arXiv},
  eprint={2310.08560},
  url={https://arxiv.org/abs/2310.08560},
  note={arXiv preprint; Accessed: 2026-04-12}
}

@article{zhang2024survey,
  title={A Survey on the Memory Mechanism of Large Language Model based Agents},
  author={Zeyu Zhang and Xiaohe Bo and Chen Ma and Rui Li and Xu Chen and Quanyu Dai and Jieming Zhu and Zhenhua Dong and Ji-Rong Wen},
  journal={arXiv preprint arXiv:2404.13501},
  year={2024},
  doi={10.48550/ARXIV.2404.13501},
  eprinttype={arXiv},
  eprint={2404.13501},
  url={https://arxiv.org/abs/2404.13501},
  note={arXiv preprint; Accessed: 2026-04-12}
}

@article{mmpoisonrag2025,
  title={MM-PoisonRAG: Disrupting Multimodal RAG with Local and Global Poisoning Attacks},
  author={Hyeonjeong Ha and Qiusi Zhan and Jeonghwan Kim and Dimitrios Bralios and Saikrishna Sanniboina and Nanyun Peng and Kai-Wei Chang and Daniel Kang and Heng Ji},
  journal={arXiv preprint arXiv:2502.17832},
  year={2025},
  doi={10.48550/ARXIV.2502.17832},
  eprinttype={arXiv},
  eprint={2502.17832},
  url={https://arxiv.org/abs/2502.17832},
  note={arXiv preprint; Accessed: 2026-04-12}
}

@article{onepic2025,
  title={One Pic is All it Takes: Poisoning Visual Document Retrieval Augmented Generation with a Single Image},
  author={Ezzeldin Shereen and Dan Ristea and Shae McFadden and Burak Hasircioglu and Vasilios Mavroudis and Chris Hicks},
  journal={arXiv preprint arXiv:2504.02132},
  year={2025},
  doi={10.48550/ARXIV.2504.02132},
  eprinttype={arXiv},
  eprint={2504.02132},
  url={https://arxiv.org/abs/2504.02132},
  note={arXiv preprint; Accessed: 2026-04-12}
}

@inproceedings{chen2024agentpoison,
  title={{AgentPoison}: Red-teaming {LLM} Agents via Poisoning Memory or Knowledge Bases},
  author={Chen, Zhaorun and Xiang, Zhen and Xiao, Chaowei and Song, Dawn and Li, Bo},
  booktitle={Advances in Neural Information Processing Systems},
  volume={37},
  year={2024},
  doi={10.48550/ARXIV.2407.12784},
  eprinttype={arXiv},
  eprint={2407.12784},
  url={https://papers.nips.cc/paper_files/paper/2024/hash/eb113910e9c3f6242541c1652e30dfd6-Abstract-Conference.html},
  note={NeurIPS proceedings page; Accessed: 2026-04-12}
}

@article{zou2024adversarial,
  title={Dissecting Adversarial Robustness of Multimodal LM Agents},
  author={Chen Henry Wu and Rishi Shah and Jing Yu Koh and Ruslan Salakhutdinov and Daniel Fried and Aditi Raghunathan},
  journal={arXiv preprint arXiv:2406.12814},
  year={2024},
  doi={10.48550/ARXIV.2406.12814},
  eprinttype={arXiv},
  eprint={2406.12814},
  url={https://arxiv.org/abs/2406.12814},
  note={arXiv preprint; Accessed: 2026-04-12}
}

@article{zhao2024revisiting,
  title={Revisiting the Adversarial Robustness of Vision Language Models: a Multimodal Perspective},
  author={Zhou, Wanqi and Bai, Shuanghao and Mandic, Danilo P. and Zhao, Qibin and Chen, Badong},
  journal={arXiv preprint arXiv:2404.19287},
  year={2024},
  doi={10.48550/ARXIV.2404.19287},
  eprinttype={arXiv},
  eprint={2404.19287},
  url={https://arxiv.org/abs/2404.19287},
  note={arXiv preprint; Accessed: 2026-04-12}
}

@article{shayegani2024jailbreak,
  title={Jailbreak in Pieces: Compositional Adversarial Attacks on Multi-Modal Language Models},
  author={Shayegani, Erfan and Dong, Yue and Abu-Ghazaleh, Nael},
  journal={arXiv preprint arXiv:2307.14539},
  year={2023},
  doi={10.48550/ARXIV.2307.14539},
  eprinttype={arXiv},
  eprint={2307.14539},
  url={https://arxiv.org/abs/2307.14539},
  note={arXiv preprint; Accessed: 2026-04-12}
}

@inproceedings{nie2022diffpure,
  title={Diffusion Models for Adversarial Purification},
  author={Nie, Weili and Guo, Brandon and Huang, Yujia and Xiao, Chaowei and Vahdat, Arash and Anandkumar, Anima},
  booktitle={International Conference on Machine Learning},
  pages={16805--16827},
  year={2022},
  organization={PMLR},
  doi={10.48550/ARXIV.2205.07460},
  eprinttype={arXiv},
  eprint={2205.07460},
  url={https://proceedings.mlr.press/v162/nie22a.html},
  note={ICML proceedings page; Accessed: 2026-04-12}
}

@book{kahneman2011thinking,
  title={Thinking, Fast and Slow},
  author={Kahneman, Daniel},
  year={2011},
  publisher={Farrar, Straus and Giroux},
  isbn={9780374533557},
  url={https://us.macmillan.com/books/9780374533557/thinkingfastandslow},
  note={Publisher book page; Accessed: 2026-04-12}
}

@misc{bengio2019system2,
  title={From System 1 Deep Learning to System 2 Deep Learning},
  author={Bengio, Yoshua},
  howpublished={Invited talk at NeurIPS 2019},
  year={2019},
  url={https://neurips.cc/virtual/2019/invited-talk/15488},
  note={NeurIPS invited talk page and SlidesLive recording; Accessed: 2026-04-12}
}

@article{liu2024persistent,
  title={MemoryGraft: Persistent Compromise of LLM Agents via Poisoned Experience Retrieval},
  author={Saksham Sahai Srivastava and Haoyu He},
  journal={arXiv preprint arXiv:2512.16962},
  year={2025},
  doi={10.48550/ARXIV.2512.16962},
  eprinttype={arXiv},
  eprint={2512.16962},
  url={https://arxiv.org/abs/2512.16962},
  note={arXiv preprint; Accessed: 2026-04-12}
}

@article{goodfellow2015explaining,
  title={Explaining and Harnessing Adversarial Examples},
  author={Goodfellow, Ian J and Shlens, Jonathon and Szegedy, Christian},
  journal={arXiv preprint arXiv:1412.6572},
  year={2015},
  doi={10.48550/ARXIV.1412.6572},
  eprinttype={arXiv},
  eprint={1412.6572},
  url={https://arxiv.org/abs/1412.6572},
  note={arXiv preprint; Accessed: 2026-04-12}
}

@inproceedings{madry2018towards,
  title={Towards Deep Learning Models Resistant to Adversarial Attacks},
  author={Madry, Aleksander and Makelov, Aleksandar and Schmidt, Ludwig and Tsipras, Dimitris and Vladu, Adrian},
  booktitle={International Conference on Learning Representations},
  year={2018},
  doi={10.48550/ARXIV.1706.06083},
  eprinttype={arXiv},
  eprint={1706.06083},
  url={https://arxiv.org/abs/1706.06083},
  note={ICLR OpenReview record: https://openreview.net/forum?id=rJzIBfZAb}
}

@misc{owasp2025agentic,
  title={{OWASP} Top 10 for Agentic Applications (2026)},
  author={{OWASP GenAI Security Project}},
  howpublished={OWASP GenAI Security Project},
  url={https://genai.owasp.org/resource/owasp-top-10-for-agentic-applications-for-2026/},
  year={2025},
  note={OWASP resource page; Published: December 9, 2025; Accessed: 2026-04-12}
}

@article{xu2025minja,
  title={Memory Injection Attacks on LLM Agents via Query-Only Interaction},
  author={Shen Dong and Shaochen Xu and Pengfei He and Yige Li and Jiliang Tang and Tianming Liu and Hui Liu and Zhen Xiang},
  journal={arXiv preprint arXiv:2503.03704},
  year={2025},
  doi={10.48550/ARXIV.2503.03704},
  eprinttype={arXiv},
  eprint={2503.03704},
  url={https://arxiv.org/abs/2503.03704},
  note={arXiv preprint; Accessed: 2026-04-12}
}

@article{perez2023hackaprompt,
  title={Ignore This Title and {HackAPrompt}: Exposing Systemic Vulnerabilities of {LLMs} through a Global Scale Prompt Hacking Competition},
  author={Schulhoff, Sander and Pinto, Jeremy and Khan, Anaum and Bouchard, Louis-Francois and Si, Chenglei and Anati, Svetlina and Tagliabue, Valen and Kost, Anson Liu and Carnahan, Christopher and Boyd-Graber, Jordan},
  journal={arXiv preprint arXiv:2311.16119},
  year={2023},
  doi={10.48550/ARXIV.2311.16119},
  eprinttype={arXiv},
  eprint={2311.16119},
  url={https://arxiv.org/abs/2311.16119},
  note={arXiv preprint; Accessed: 2026-04-12}
}

@article{llama3,
  title={The Llama 3 Herd of Models},
  author={Aaron Grattafiori and Abhimanyu Dubey and Abhinav Jauhri and Abhinav Pandey and Abhishek Kadian and Ahmad Al-Dahle and Aiesha Letman and Akhil Mathur and Alan Schelten and Alex Vaughan and Amy Yang and Angela Fan and Anirudh Goyal and Anthony Hartshorn and Aobo Yang and Archi Mitra and Archie Sravankumar and Artem Korenev and Arthur Hinsvark and Arun Rao and Aston Zhang and Aurelien Rodriguez and Austen Gregerson and Ava Spataru and Baptiste Roziere and Bethany Biron and Binh Tang and Bobbie Chern and Charlotte Caucheteux and Chaya Nayak and Chloe Bi and Chris Marra and Chris McConnell and Christian Keller and Christophe Touret and Chunyang Wu and Corinne Wong and Cristian Canton Ferrer and Cyrus Nikolaidis and Damien Allonsius and Daniel Song and Danielle Pintz and Danny Livshits and Danny Wyatt and David Esiobu and Dhruv Choudhary and Dhruv Mahajan and Diego Garcia-Olano and Diego Perino and Dieuwke Hupkes and Egor Lakomkin and Ehab AlBadawy and Elina Lobanova and Emily Dinan and Eric Michael Smith and Filip Radenovic and Francisco Guzmán and Frank Zhang and Gabriel Synnaeve and Gabrielle Lee and Georgia Lewis Anderson and Govind Thattai and Graeme Nail and Gregoire Mialon and Guan Pang and Guillem Cucurell and Hailey Nguyen and Hannah Korevaar and Hu Xu and Hugo Touvron and Iliyan Zarov and Imanol Arrieta Ibarra and Isabel Kloumann and Ishan Misra and Ivan Evtimov and Jack Zhang and Jade Copet and Jaewon Lee and Jan Geffert and Jana Vranes and Jason Park and Jay Mahadeokar and Jeet Shah and Jelmer van der Linde and Jennifer Billock and Jenny Hong and Jenya Lee and Jeremy Fu and Jianfeng Chi and Jianyu Huang and Jiawen Liu and Jie Wang and Jiecao Yu and Joanna Bitton and Joe Spisak and Jongsoo Park and Joseph Rocca and Joshua Johnstun and Joshua Saxe and Junteng Jia and Kalyan Vasuden Alwala and Karthik Prasad and Kartikeya Upasani and Kate Plawiak and Ke Li and Kenneth Heafield and Kevin Stone and Khalid El-Arini and Krithika Iyer and Kshitiz Malik and Kuenley Chiu and Kunal Bhalla and Kushal Lakhotia and Lauren Rantala-Yeary and Laurens van der Maaten and Lawrence Chen and Liang Tan and Liz Jenkins and Louis Martin and Lovish Madaan and Lubo Malo and Lukas Blecher and Lukas Landzaat and Luke de Oliveira and Madeline Muzzi and Mahesh Pasupuleti and Mannat Singh and Manohar Paluri and Marcin Kardas and Maria Tsimpoukelli and Mathew Oldham and Mathieu Rita and Maya Pavlova and Melanie Kambadur and Mike Lewis and Min Si and Mitesh Kumar Singh and Mona Hassan and Naman Goyal and Narjes Torabi and Nikolay Bashlykov and Nikolay Bogoychev and Niladri Chatterji and Ning Zhang and Olivier Duchenne and Onur Çelebi and Patrick Alrassy and Pengchuan Zhang and Pengwei Li and Petar Vasic and Peter Weng and Prajjwal Bhargava and Pratik Dubal and Praveen Krishnan and Punit Singh Koura and Puxin Xu and Qing He and Qingxiao Dong and Ragavan Srinivasan and Raj Ganapathy and Ramon Calderer and Ricardo Silveira Cabral and Robert Stojnic and Roberta Raileanu and Rohan Maheswari and Rohit Girdhar and Rohit Patel and Romain Sauvestre and Ronnie Polidoro and Roshan Sumbaly and Ross Taylor and Ruan Silva and Rui Hou and Rui Wang and Saghar Hosseini and Sahana Chennabasappa and Sanjay Singh and Sean Bell and Seohyun Sonia Kim and Sergey Edunov and Shaoliang Nie and Sharan Narang and Sharath Raparthy and Sheng Shen and Shengye Wan and Shruti Bhosale and Shun Zhang and Simon Vandenhende and Soumya Batra and Spencer Whitman and Sten Sootla and Stephane Collot and Suchin Gururangan and Sydney Borodinsky and Tamar Herman and Tara Fowler and Tarek Sheasha and Thomas Georgiou and Thomas Scialom and Tobias Speckbacher and Todor Mihaylov and Tong Xiao and Ujjwal Karn and Vedanuj Goswami and Vibhor Gupta and Vignesh Ramanathan and Viktor Kerkez and Vincent Gonguet and Virginie Do and Vish Vogeti and Vítor Albiero and Vladan Petrovic and Weiwei Chu and Wenhan Xiong and Wenyin Fu and Whitney Meers and Xavier Martinet and Xiaodong Wang and Xiaofang Wang and Xiaoqing Ellen Tan and Xide Xia and Xinfeng Xie and Xuchao Jia and Xuewei Wang and Yaelle Goldschlag and Yashesh Gaur and Yasmine Babaei and Yi Wen and Yiwen Song and Yuchen Zhang and Yue Li and Yuning Mao and Zacharie Delpierre Coudert and Zheng Yan and Zhengxing Chen and Zoe Papakipos and Aaditya Singh and Aayushi Srivastava and Abha Jain and Adam Kelsey and Adam Shajnfeld and Adithya Gangidi and Adolfo Victoria and Ahuva Goldstand and Ajay Menon and Ajay Sharma and Alex Boesenberg and Alexei Baevski and Allie Feinstein and Amanda Kallet and Amit Sangani and Amos Teo and Anam Yunus and Andrei Lupu and Andres Alvarado and Andrew Caples and Andrew Gu and Andrew Ho and Andrew Poulton and Andrew Ryan and Ankit Ramchandani and Annie Dong and Annie Franco and Anuj Goyal and Aparajita Saraf and Arkabandhu Chowdhury and Ashley Gabriel and Ashwin Bharambe and Assaf Eisenman and Azadeh Yazdan and Beau James and Ben Maurer and Benjamin Leonhardi and Bernie Huang and Beth Loyd and Beto De Paola and Bhargavi Paranjape and Bing Liu and Bo Wu and Boyu Ni and Braden Hancock and Bram Wasti and Brandon Spence and Brani Stojkovic and Brian Gamido and Britt Montalvo and Carl Parker and Carly Burton and Catalina Mejia and Ce Liu and Changhan Wang and Changkyu Kim and Chao Zhou and Chester Hu and Ching-Hsiang Chu and Chris Cai and Chris Tindal and Christoph Feichtenhofer and Cynthia Gao and Damon Civin and Dana Beaty and Daniel Kreymer and Daniel Li and David Adkins and David Xu and Davide Testuggine and Delia David and Devi Parikh and Diana Liskovich and Didem Foss and Dingkang Wang and Duc Le and Dustin Holland and Edward Dowling and Eissa Jamil and Elaine Montgomery and Eleonora Presani and Emily Hahn and Emily Wood and Eric-Tuan Le and Erik Brinkman and Esteban Arcaute and Evan Dunbar and Evan Smothers and Fei Sun and Felix Kreuk and Feng Tian and Filippos Kokkinos and Firat Ozgenel and Francesco Caggioni and Frank Kanayet and Frank Seide and Gabriela Medina Florez and Gabriella Schwarz and Gada Badeer and Georgia Swee and Gil Halpern and Grant Herman and Grigory Sizov and Guangyi and Zhang and Guna Lakshminarayanan and Hakan Inan and Hamid Shojanazeri and Han Zou and Hannah Wang and Hanwen Zha and Haroun Habeeb and Harrison Rudolph and Helen Suk and Henry Aspegren and Hunter Goldman and Hongyuan Zhan and Ibrahim Damlaj and Igor Molybog and Igor Tufanov and Ilias Leontiadis and Irina-Elena Veliche and Itai Gat and Jake Weissman and James Geboski and James Kohli and Janice Lam and Japhet Asher and Jean-Baptiste Gaya and Jeff Marcus and Jeff Tang and Jennifer Chan and Jenny Zhen and Jeremy Reizenstein and Jeremy Teboul and Jessica Zhong and Jian Jin and Jingyi Yang and Joe Cummings and Jon Carvill and Jon Shepard and Jonathan McPhie and Jonathan Torres and Josh Ginsburg and Junjie Wang and Kai Wu and Kam Hou U and Karan Saxena and Kartikay Khandelwal and Katayoun Zand and Kathy Matosich and Kaushik Veeraraghavan and Kelly Michelena and Keqian Li and Kiran Jagadeesh and Kun Huang and Kunal Chawla and Kyle Huang and Lailin Chen and Lakshya Garg and Lavender A and Leandro Silva and Lee Bell and Lei Zhang and Liangpeng Guo and Licheng Yu and Liron Moshkovich and Luca Wehrstedt and Madian Khabsa and Manav Avalani and Manish Bhatt and Martynas Mankus and Matan Hasson and Matthew Lennie and Matthias Reso and Maxim Groshev and Maxim Naumov and Maya Lathi and Meghan Keneally and Miao Liu and Michael L. Seltzer and Michal Valko and Michelle Restrepo and Mihir Patel and Mik Vyatskov and Mikayel Samvelyan and Mike Clark and Mike Macey and Mike Wang and Miquel Jubert Hermoso and Mo Metanat and Mohammad Rastegari and Munish Bansal and Nandhini Santhanam and Natascha Parks and Natasha White and Navyata Bawa and Nayan Singhal and Nick Egebo and Nicolas Usunier and Nikhil Mehta and Nikolay Pavlovich Laptev and Ning Dong and Norman Cheng and Oleg Chernoguz and Olivia Hart and Omkar Salpekar and Ozlem Kalinli and Parkin Kent and Parth Parekh and Paul Saab and Pavan Balaji and Pedro Rittner and Philip Bontrager and Pierre Roux and Piotr Dollar and Polina Zvyagina and Prashant Ratanchandani and Pritish Yuvraj and Qian Liang and Rachad Alao and Rachel Rodriguez and Rafi Ayub and Raghotham Murthy and Raghu Nayani and Rahul Mitra and Rangaprabhu Parthasarathy and Raymond Li and Rebekkah Hogan and Robin Battey and Rocky Wang and Russ Howes and Ruty Rinott and Sachin Mehta and Sachin Siby and Sai Jayesh Bondu and Samyak Datta and Sara Chugh and Sara Hunt and Sargun Dhillon and Sasha Sidorov and Satadru Pan and Saurabh Mahajan and Saurabh Verma and Seiji Yamamoto and Sharadh Ramaswamy and Shaun Lindsay and Shaun Lindsay and Sheng Feng and Shenghao Lin and Shengxin Cindy Zha and Shishir Patil and Shiva Shankar and Shuqiang Zhang and Shuqiang Zhang and Sinong Wang and Sneha Agarwal and Soji Sajuyigbe and Soumith Chintala and Stephanie Max and Stephen Chen and Steve Kehoe and Steve Satterfield and Sudarshan Govindaprasad and Sumit Gupta and Summer Deng and Sungmin Cho and Sunny Virk and Suraj Subramanian and Sy Choudhury and Sydney Goldman and Tal Remez and Tamar Glaser and Tamara Best and Thilo Koehler and Thomas Robinson and Tianhe Li and Tianjun Zhang and Tim Matthews and Timothy Chou and Tzook Shaked and Varun Vontimitta and Victoria Ajayi and Victoria Montanez and Vijai Mohan and Vinay Satish Kumar and Vishal Mangla and Vlad Ionescu and Vlad Poenaru and Vlad Tiberiu Mihailescu and Vladimir Ivanov and Wei Li and Wenchen Wang and Wenwen Jiang and Wes Bouaziz and Will Constable and Xiaocheng Tang and Xiaojian Wu and Xiaolan Wang and Xilun Wu and Xinbo Gao and Yaniv Kleinman and Yanjun Chen and Ye Hu and Ye Jia and Ye Qi and Yenda Li and Yilin Zhang and Ying Zhang and Yossi Adi and Youngjin Nam and Yu and Wang and Yu Zhao and Yuchen Hao and Yundi Qian and Yunlu Li and Yuzi He and Zach Rait and Zachary DeVito and Zef Rosnbrick and Zhaoduo Wen and Zhenyu Yang and Zhiwei Zhao and Zhiyu Ma},
  journal={arXiv preprint arXiv:2407.21783},
  year={2024},
  doi={10.48550/ARXIV.2407.21783},
  eprinttype={arXiv},
  eprint={2407.21783},
  url={https://arxiv.org/abs/2407.21783},
  note={arXiv preprint; Accessed: 2026-04-12}
}

@techreport{openai2024gpt4v,
  title={{GPT-4V(ision)} System Card},
  author={{OpenAI}},
  institution={OpenAI},
  year={2023},
  month={sep},
  url={https://cdn.openai.com/papers/GPTV_System_Card.pdf},
  note={OpenAI system card (PDF); Published: September 2023; Accessed: 2026-04-12}
}

@article{faiss,
  title={Billion-Scale Similarity Search with {GPUs}},
  author={Johnson, Jeff and Douze, Matthijs and J{\'e}gou, Herv{\'e}},
  journal={IEEE Transactions on Big Data},
  volume={7},
  number={3},
  pages={535--547},
  year={2021},
  doi={10.1109/TBDATA.2019.2921572},
  url={https://ieeexplore.ieee.org/document/8733051},
  note={IEEE Xplore landing page; Accessed: 2026-04-12}
}

@article{proactivedefense2025,
  title={A-MemGuard: A Proactive Defense Framework for LLM-Based Agent Memory},
  author={Qianshan Wei and Tengchao Yang and Yaochen Wang and Xinfeng Li and Lijun Li and Zhenfei Yin and Yi Zhan and Thorsten Holz and Zhiqiang Lin and XiaoFeng Wang},
  journal={arXiv preprint arXiv:2510.02373},
  year={2025},
  doi={10.48550/ARXIV.2510.02373},
  eprinttype={arXiv},
  eprint={2510.02373},
  url={https://arxiv.org/abs/2510.02373},
  note={arXiv preprint; Accessed: 2026-04-12}
}

@article{reimers2019sentence,
  title={Sentence-{BERT}: Sentence Embeddings using Siamese {BERT}-Networks},
  author={Reimers, Nils and Gurevych, Iryna},
  journal={arXiv preprint arXiv:1908.10084},
  year={2019},
  doi={10.48550/ARXIV.1908.10084},
  eprinttype={arXiv},
  eprint={1908.10084},
  url={https://arxiv.org/abs/1908.10084},
  note={arXiv preprint; Accessed: 2026-04-12}
}

@inproceedings{tramer2020adaptive,
  title={On Adaptive Attacks to Adversarial Example Defenses},
  author={Tramer, Florian and Carlini, Nicholas and Brendel, Wieland and Madry, Aleksander},
  booktitle={Advances in Neural Information Processing Systems},
  volume={33},
  pages={1633--1645},
  year={2020},
  doi={10.48550/ARXIV.2002.08347},
  eprinttype={arXiv},
  eprint={2002.08347},
  url={https://proceedings.neurips.cc/paper/2020/hash/11f38f8ecd71867b42433548d1078e38-Abstract.html},
  note={NeurIPS proceedings page; Accessed: 2026-04-12}
}

@inproceedings{croce2022evaluating,
  title={Evaluating the Adversarial Robustness of Adaptive Test-time Defenses},
  author={Croce, Francesco and Gowal, Sven and Brunner, Thomas and Shelhamer, Evan and Hein, Matthias and Cemgil, Taylan},
  booktitle={International Conference on Machine Learning},
  pages={4421--4435},
  year={2022},
  organization={PMLR},
  doi={10.48550/ARXIV.2202.13711},
  eprinttype={arXiv},
  eprint={2202.13711},
  url={https://proceedings.mlr.press/v162/croce22a.html},
  note={ICML proceedings page; Accessed: 2026-04-12}
}

@article{xiang2024certifiably,
  title={Certifiably Robust {RAG} against Retrieval Corruption},
  author={Xiang, Chong and Wu, Tong and Zhong, Zexuan and Wagner, David and Chen, Danqi and Mittal, Prateek},
  journal={arXiv preprint arXiv:2405.15556},
  year={2024},
  doi={10.48550/ARXIV.2405.15556},
  eprinttype={arXiv},
  eprint={2405.15556},
  url={https://arxiv.org/abs/2405.15556},
  note={arXiv preprint; Accessed: 2026-04-12}
}

@article{zou2024poisonedrag,
  title={PoisonedRAG: Knowledge Corruption Attacks to Retrieval-Augmented Generation of Large Language Models},
  author={Wei Zou and Runpeng Geng and Binghui Wang and Jinyuan Jia},
  journal={arXiv preprint arXiv:2402.07867},
  year={2024},
  doi={10.48550/ARXIV.2402.07867},
  eprinttype={arXiv},
  eprint={2402.07867},
  url={https://arxiv.org/abs/2402.07867},
  note={arXiv preprint; Accessed: 2026-04-12}
}

@article{schlarmann2024robust,
  title={Robust {CLIP}: Unsupervised Adversarial Fine-Tuning of Vision Embeddings for Robust Large Vision-Language Models},
  author={Schlarmann, Christian and Singh, Naman Deep and Croce, Francesco and Hein, Matthias},
  journal={arXiv preprint arXiv:2402.12336},
  year={2024},
  doi={10.48550/ARXIV.2402.12336},
  eprinttype={arXiv},
  eprint={2402.12336},
  url={https://arxiv.org/abs/2402.12336},
  note={arXiv preprint; Accessed: 2026-04-12}
}

@article{saferag2025,
  title={SafeRAG: Benchmarking Security in Retrieval-Augmented Generation of Large Language Model},
  author={Xun Liang and Simin Niu and Zhiyu Li and Sensen Zhang and Hanyu Wang and Feiyu Xiong and Jason Zhaoxin Fan and Bo Tang and Shichao Song and Mengwei Wang and Jiawei Yang},
  journal={arXiv preprint arXiv:2501.18636},
  year={2025},
  doi={10.48550/ARXIV.2501.18636},
  eprinttype={arXiv},
  eprint={2501.18636},
  url={https://arxiv.org/abs/2501.18636},
  note={arXiv preprint; Accessed: 2026-04-12}
}

@article{ragbench2024,
  title={{RAGBench}: Explainable Benchmark for Retrieval-Augmented Generation Systems},
  author={Friel, Robert and Belyi, Masha and Sanyal, Atindriyo},
  journal={arXiv preprint arXiv:2407.11005},
  year={2024},
  doi={10.48550/ARXIV.2407.11005},
  eprinttype={arXiv},
  eprint={2407.11005},
  url={https://arxiv.org/abs/2407.11005},
  note={arXiv preprint; Accessed: 2026-04-12}
}

@article{diffbreak2024,
  title={DiffBreak: Is Diffusion-Based Purification Robust?},
  author={Andre Kassis and Urs Hengartner and Yaoliang Yu},
  journal={arXiv preprint arXiv:2411.16598},
  year={2024},
  doi={10.48550/ARXIV.2411.16598},
  eprinttype={arXiv},
  eprint={2411.16598},
  url={https://arxiv.org/abs/2411.16598},
  note={arXiv preprint; Accessed: 2026-04-12}
}

@article{simcse2021,
  title={{SimCSE}: Simple Contrastive Learning of Sentence Embeddings},
  author={Gao, Tianyu and Yao, Xingcheng and Chen, Danqi},
  journal={Proceedings of the 2021 Conference on Empirical Methods in Natural Language Processing},
  pages={6894--6910},
  year={2021},
  doi={10.18653/v1/2021.emnlp-main.552},
  url={https://aclanthology.org/2021.emnlp-main.552/},
  note={ACL Anthology entry; Accessed: 2026-04-12}
}

@article{e5embedding2024,
  title={Text Embeddings by Weakly-Supervised Contrastive Pre-training},
  author={Liang Wang and Nan Yang and Xiaolong Huang and Binxing Jiao and Linjun Yang and Daxin Jiang and Rangan Majumder and Furu Wei},
  journal={arXiv preprint arXiv:2212.03533},
  year={2022},
  doi={10.48550/ARXIV.2212.03533},
  eprinttype={arXiv},
  eprint={2212.03533},
  url={https://arxiv.org/abs/2212.03533},
  note={arXiv preprint; Accessed: 2026-04-12}
}

@article{ilyas2018black,
  title={Black-box Adversarial Attacks with Limited Queries and Information},
  author={Andrew Ilyas and Logan Engstrom and Anish Athalye and Jessy Lin},
  journal={arXiv preprint arXiv:1804.08598},
  year={2018},
  doi={10.48550/ARXIV.1804.08598},
  eprinttype={arXiv},
  eprint={1804.08598},
  url={https://arxiv.org/abs/1804.08598},
  note={arXiv preprint; Accessed: 2026-04-12}
}

@inproceedings{brendel2018decision,
  title={Decision-Based Adversarial Attacks: Reliable Attacks Against Black-box Machine Learning Models},
  author={Brendel, Wieland and Rauber, Jonas and Bethge, Matthias},
  booktitle={International Conference on Learning Representations},
  year={2018},
  doi={10.48550/ARXIV.1712.04248},
  eprinttype={arXiv},
  eprint={1712.04248},
  url={https://arxiv.org/abs/1712.04248},
  note={ICLR OpenReview record: https://openreview.net/forum?id=SyZI0GWCZ}
}

@article{liang2024badclip,
  title={{BadCLIP}: Dual-Embedding Guided Backdoor Attack on Multimodal Contrastive Learning},
  author={Liang, Siyuan and Zhu, Mingli and Liu, Aishan and Wu, Baoyuan and Cao, Xiaochun and Chang, Ee-Chien},
  journal={arXiv preprint arXiv:2311.12075},
  year={2023},
  doi={10.48550/ARXIV.2311.12075},
  eprinttype={arXiv},
  eprint={2311.12075},
  url={https://arxiv.org/abs/2311.12075},
  note={arXiv preprint; Accessed: 2026-04-12}
}

@article{wang2024adbm,
  title={{ADBM}: Adversarial diffusion bridge model for reliable adversarial purification},
  author={Li, Xiao and Sun, Wenxuan and Chen, Huanran and Li, Qiongxiu and Liu, Yining and He, Yingzhe and Shi, Jie and Hu, Xiaolin},
  journal={arXiv preprint arXiv:2408.00315},
  year={2024},
  doi={10.48550/ARXIV.2408.00315},
  eprinttype={arXiv},
  eprint={2408.00315},
  url={https://arxiv.org/abs/2408.00315},
  note={arXiv preprint; Accessed: 2026-04-12}
}

@article{chen2024agdm,
  title={Adversarial Guided Diffusion Models for Adversarial Purification},
  author={Lin, Guang and Tao, Zerui and Zhang, Jianhai and Tanaka, Toshihisa and Zhao, Qibin},
  journal={arXiv preprint arXiv:2403.16067},
  year={2024},
  doi={10.48550/ARXIV.2403.16067},
  eprinttype={arXiv},
  eprint={2403.16067},
  url={https://arxiv.org/abs/2403.16067},
  note={arXiv preprint; Accessed: 2026-04-12}
}

@article{zhang2024codefend,
  title={{CoDefend}: Cross-Modal Collaborative Defense via Diffusion Purification and Prompt Optimization},
  author={Zhu, Fengling and Liu, Boshi and Hua, Jingyu and Zhong, Sheng},
  journal={arXiv preprint arXiv:2510.11096},
  year={2025},
  doi={10.48550/ARXIV.2510.11096},
  eprinttype={arXiv},
  eprint={2510.11096},
  url={https://arxiv.org/abs/2510.11096},
  note={arXiv preprint; Accessed: 2026-04-12}
}

@article{liu2024crossfire,
  title={Adversarial Attacks to Multi-Modal Models},
  author={Dou, Zhihao and Hu, Xin and Yang, Haibo and Liu, Zhuqing and Fang, Minghong},
  journal={arXiv preprint arXiv:2409.06793},
  year={2024},
  doi={10.48550/ARXIV.2409.06793},
  eprinttype={arXiv},
  eprint={2409.06793},
  url={https://arxiv.org/abs/2409.06793},
  note={Introduces CrossFire attack}
}

@book{pearl2009causality,
  title={Causality: Models, Reasoning, and Inference},
  author={Pearl, Judea},
  year={2009},
  publisher={Cambridge University Press},
  edition={2nd},
  doi={10.1017/CBO9780511803161},
  url={https://doi.org/10.1017/CBO9780511803161},
  note={Book DOI landing page; Accessed: 2026-04-12}
}

@inproceedings{wang2019neural,
  title={Neural Cleanse: Identifying and Mitigating Backdoor Attacks in Neural Networks},
  author={Wang, Bolun and Yao, Yuanshun and Shan, Shawn and Li, Huiying and Viswanath, Bimal and Zheng, Haitao and Zhao, Ben Y},
  booktitle={2019 IEEE Symposium on Security and Privacy (SP)},
  pages={707--723},
  year={2019},
  organization={IEEE},
  doi={10.1109/SP.2019.00031},
  url={https://ieeexplore.ieee.org/document/8835365},
  note={IEEE Xplore landing page; Accessed: 2026-04-12}
}

@inproceedings{gao2019strip,
  title={{STRIP}: A Defence Against Trojan Attacks on Deep Neural Networks},
  author={Gao, Yansong and Xu, Change and Wang, Derui and Chen, Shiping and Ranasinghe, Damith C and Nepal, Surya},
  booktitle={Proceedings of the 35th Annual Computer Security Applications Conference},
  pages={113--125},
  year={2019},
  doi={10.1145/3359789.3359790},
  eprinttype={arXiv},
  eprint={1902.06531},
  url={https://dl.acm.org/doi/10.1145/3359789.3359790},
  note={ACM Digital Library landing page; Accessed: 2026-04-12}
}

@inproceedings{tran2018spectral,
  title={Spectral Signatures in Backdoor Attacks},
  author={Tran, Brandon and Li, Jerry and Madry, Aleksander},
  booktitle={Advances in Neural Information Processing Systems},
  volume={31},
  year={2018},
  doi={10.48550/ARXIV.1811.00636},
  eprinttype={arXiv},
  eprint={1811.00636},
  url={https://proceedings.neurips.cc/paper/2018/hash/280cf18baf4311c92aa5a042336587d3-Abstract.html},
  note={NeurIPS proceedings page; Accessed: 2026-04-12}
}

@article{trustrag2024,
  title={TrustRAG: An Information Assistant with Retrieval Augmented Generation},
  author={Yixing Fan and Qiang Yan and Wenshan Wang and Jiafeng Guo and Ruqing Zhang and Xueqi Cheng},
  journal={arXiv preprint arXiv:2502.13719},
  year={2025},
  doi={10.48550/ARXIV.2502.13719},
  eprinttype={arXiv},
  eprint={2502.13719},
  url={https://arxiv.org/abs/2502.13719},
  note={arXiv preprint; Accessed: 2026-04-12}
}

@inproceedings{peng2025llmagent,
  title={A Survey on {LLM}-powered Agents for Recommender Systems},
  author={Peng, Qiyao and Liu, Hongtao and Huang, Hua and Yang, Jian and Yang, Qing and Shao, Minglai},
  booktitle={Findings of the Association for Computational Linguistics: EMNLP 2025},
  pages={11574--11583},
  year={2025},
  address={Suzhou, China},
  doi={10.18653/v1/2025.findings-emnlp.620},
  url={https://aclanthology.org/2025.findings-emnlp.620/},
  note={ACL Anthology entry; Accessed: 2026-04-12}
}

@inproceedings{wang2024recai,
  title={{RecAI}: Leveraging Large Language Models for Next-Generation Recommender Systems},
  author={Lian, Jianxun and Lei, Yuxuan and Huang, Xu and Yao, Jing and Xu, Wei and Xie, Xing},
  booktitle={Companion Proceedings of the ACM Web Conference 2024},
  pages={1--4},
  year={2024},
  organization={ACM},
  doi={10.1145/3589335.3651242},
  url={https://dl.acm.org/doi/10.1145/3589335.3651242},
  note={ACM Digital Library landing page; Accessed: 2026-04-12}
}

@inproceedings{wang2024ragbestpractices,
  title={Searching for Best Practices in Retrieval-Augmented Generation},
  author={Wang, Xiaohua and Wang, Zhenghua and Gao, Xuan and Zhang, Feiran and Wu, Yixin and Xu, Zhibo and Shi, Tianyuan and Wang, Zhengyuan and Li, Shizheng and Qian, Qi and Yin, Ruicheng and Lv, Changze and Zheng, Xiaoqing and Huang, Xuanjing},
  booktitle={Proceedings of the 2024 Conference on Empirical Methods in Natural Language Processing},
  pages={17716--17736},
  year={2024},
  address={Miami, Florida, USA},
  doi={10.18653/v1/2024.emnlp-main.981},
  url={https://aclanthology.org/2024.emnlp-main.981/},
  note={ACL Anthology entry; Accessed: 2026-04-12}
}

@inproceedings{peris2024reducing,
  title={Reducing Hallucination in Structured Outputs via Retrieval-Augmented Generation},
  author={Ayala, Orlando and Bechard, Patrice},
  booktitle={Proceedings of the 2024 Conference of the North American Chapter of the Association for Computational Linguistics: Human Language Technologies (Volume 6: Industry Track)},
  pages={228--238},
  year={2024},
  address={Mexico City, Mexico},
  doi={10.18653/v1/2024.naacl-industry.19},
  url={https://aclanthology.org/2024.naacl-industry.19/},
  note={ACL Anthology entry; Accessed: 2026-04-12}
}

@inproceedings{hung2025attention,
  title={Attention Tracker: Detecting Prompt Injection Attacks in {LLMs}},
  author={Hung, Kuo-Han and Ko, Ching-Yun and Rawat, Ambrish and Chung, I-Hsin and Hsu, Winston H. and Chen, Pin-Yu},
  booktitle={Findings of the Association for Computational Linguistics: NAACL 2025},
  year={2025},
  address={Albuquerque, New Mexico},
  pages={2309--2322},
  doi={10.18653/v1/2025.findings-naacl.123},
  url={https://aclanthology.org/2025.findings-naacl.123/},
  note={ACL Anthology entry; Accessed: 2026-04-12}
}

@inproceedings{chen2025defense,
  title={Defense Against Prompt Injection Attack by Leveraging Attack Techniques},
  author={Chen, Yulin and Li, Haoran and Zheng, Zihao and Wu, Dekai and Song, Yangqiu and Hooi, Bryan},
  booktitle={Proceedings of the 63rd Annual Meeting of the Association for Computational Linguistics (Volume 1: Long Papers)},
  year={2025},
  address={Vienna, Austria},
  pages={18331--18347},
  doi={10.18653/v1/2025.acl-long.897},
  url={https://aclanthology.org/2025.acl-long.897/},
  note={ACL Anthology entry; Accessed: 2026-04-12}
}

@inproceedings{liu2025fcattack,
  title={{FC-Attack}: Jailbreaking Multimodal Large Language Models via Auto-Generated Flowcharts},
  author={Zhang, Ziyi and Sun, Zhen and Zhang, Zongmin and Guo, Jihui and He, Xinlei},
  booktitle={Findings of the Association for Computational Linguistics: EMNLP 2025},
  pages={9299--9316},
  year={2025},
  address={Suzhou, China},
  doi={10.18653/v1/2025.findings-emnlp.494},
  url={https://aclanthology.org/2025.findings-emnlp.494/},
  note={ACL Anthology entry; Accessed: 2026-04-12}
}

@article{wang2025immune,
  title={Immune: Improving Safety Against Jailbreaks in Multi-modal LLMs via Inference-Time Alignment},
  author={Soumya Suvra Ghosal and Souradip Chakraborty and Vaibhav Singh and Tianrui Guan and Mengdi Wang and Alvaro Velasquez and Ahmad Beirami and Furong Huang and Dinesh Manocha and Amrit Singh Bedi},
  journal={arXiv preprint arXiv:2411.18688},
  year={2024},
  doi={10.48550/ARXIV.2411.18688},
  eprinttype={arXiv},
  eprint={2411.18688},
  url={https://arxiv.org/abs/2411.18688},
  note={arXiv preprint; Accessed: 2026-04-12}
}

@inproceedings{li2025chainofscrutiny,
  title={Chain-of-Scrutiny: Detecting Backdoor Attacks for Large Language Models},
  author={Li, Xi and Mao, Ruofan and Zhang, Yusen and Lou, Renze and Wu, Chen and Wang, Jiaqi},
  booktitle={Findings of the Association for Computational Linguistics: ACL 2025},
  pages={7705--7727},
  year={2025},
  address={Vienna, Austria},
  doi={10.18653/v1/2025.findings-acl.401},
  url={https://aclanthology.org/2025.findings-acl.401/},
  note={ACL Anthology entry; Accessed: 2026-04-12}
}

@article{perez2025phantom,
  title={Phantom: General Backdoor Attacks on Retrieval Augmented Language Generation},
  author={Chaudhari, Harsh and Severi, Giorgio and Abascal, John and Suri, Anshuman and Jagielski, Matthew and Choquette-Choo, Christopher A. and Nasr, Milad and Nita-Rotaru, Cristina and Oprea, Alina},
  journal={CoRR},
  volume={abs/2405.20485},
  year={2024},
  doi={10.48550/ARXIV.2405.20485},
  eprinttype={arXiv},
  eprint={2405.20485},
  url={https://arxiv.org/abs/2405.20485},
  note={arXiv preprint; Accessed: 2026-04-12}
}

@article{zhang2025neurogenpoison,
  title={NeuroGenPoisoning: Neuron-Guided Attacks on Retrieval-Augmented Generation of {LLM} via Genetic Optimization of External Knowledge},
  author={Zhu, Hanyu and Fiondella, Lance and Yuan, Jiawei and Zeng, Kai and Jiao, Long},
  journal={CoRR},
  volume={abs/2510.21144},
  year={2025},
  doi={10.48550/ARXIV.2510.21144},
  eprinttype={arXiv},
  eprint={2510.21144},
  url={https://arxiv.org/abs/2510.21144},
  note={arXiv preprint; Accessed: 2026-04-12}
}

@inproceedings{wang2025tapt,
  title={{TAPT}: Test-Time Adversarial Prompt Tuning for Robust Inference in Vision-Language Models},
  author={Wang, Xin and Chen, Kai and Zhang, Jiaming and Chen, Jingjing and Ma, Xingjun},
  booktitle={Proceedings of the IEEE/CVF Conference on Computer Vision and Pattern Recognition},
  year={2025},
  url={https://openaccess.thecvf.com/content/CVPR2025/html/Wang_TAPT_Test-Time_Adversarial_Prompt_Tuning_for_Robust_Inference_in_Vision-Language_CVPR_2025_paper.html},
  note={CVPR 2025 open-access paper page; Accessed: 2026-04-12}
}

@article{kumar2025simclipplus,
  title={Securing Vision-Language Models with a Robust Encoder Against Jailbreak and Adversarial Attacks},
  author={Hossain, Md Zarif and Imteaj, Ahmed},
  journal={arXiv preprint arXiv:2409.07353},
  year={2024},
  doi={10.48550/ARXIV.2409.07353},
  eprinttype={arXiv},
  eprint={2409.07353},
  url={https://arxiv.org/abs/2409.07353},
  note={arXiv preprint; Accessed: 2026-04-12}
}

@article{hadi2025simclip,
  title={{Sim-CLIP}: Unsupervised Siamese Adversarial Fine-Tuning for Robust and Semantically-Rich Vision-Language Models},
  author={Hossain, Md Zarif and Imteaj, Ahmed},
  journal={arXiv preprint arXiv:2407.14971},
  year={2024},
  doi={10.48550/ARXIV.2407.14971},
  eprinttype={arXiv},
  eprint={2407.14971},
  url={https://arxiv.org/abs/2407.14971},
  note={arXiv preprint; Accessed: 2026-04-12}
}

@article{kiciman2024causal,
  title={Causal Reasoning and Large Language Models: Opening a New Frontier for Causality},
  author={Kiciman, Emre and Ness, Robert and Sharma, Amit and Tan, Chenhao},
  journal={Transactions on Machine Learning Research},
  year={2024},
  doi={10.48550/ARXIV.2305.00050},
  eprinttype={arXiv},
  eprint={2305.00050},
  url={https://arxiv.org/abs/2305.00050},
  note={TMLR OpenReview record: https://openreview.net/forum?id=mqoxLkX210}
}

@article{chen2025counterfactual,
  title={Counterfactual Causal Inference in Natural Language with Large Language Models},
  author={Gendron, Ga{\"e}l and Rozanec, Joze M. and Witbrock, Michael and Dobbie, Gillian},
  journal={CoRR},
  volume={abs/2410.06392},
  year={2024},
  doi={10.48550/ARXIV.2410.06392},
  eprinttype={arXiv},
  eprint={2410.06392},
  url={https://arxiv.org/abs/2410.06392},
  note={arXiv preprint; Accessed: 2026-04-12}
}

@inproceedings{ross2024faithful,
  title={Faithful Explanations of Black-box {NLP} Models Using {LLM}-generated Counterfactuals},
  author={Gat, Yair Ori and Calderon, Nitay and Feder, Amir and Chapanin, Alexander and Sharma, Amit and Reichart, Roi},
  booktitle={The Twelfth International Conference on Learning Representations},
  year={2024},
  doi={10.48550/ARXIV.2310.00603},
  eprinttype={arXiv},
  eprint={2310.00603},
  url={https://arxiv.org/abs/2310.00603},
  note={ICLR OpenReview record: https://openreview.net/forum?id=UMfcdRIotC}
}

@inproceedings{ji2025hallucination,
  title={Developing a Reliable, Fast, General-Purpose Hallucination Detection and Mitigation Service},
  author={Wang, Song and Wang, Xun and Mei, Jie and Xie, Yujia and Chen, Si-Qing and Xiong, Wayne},
  booktitle={Proceedings of the 2025 Conference of the North American Chapter of the Association for Computational Linguistics: Human Language Technologies (Volume 3: Industry Track)},
  year={2025},
  address={Albuquerque, New Mexico},
  pages={971--978},
  doi={10.18653/v1/2025.naacl-industry.72},
  url={https://aclanthology.org/2025.naacl-industry.72/},
  note={ACL Anthology entry; Accessed: 2026-04-12}
}

@article{zhang2024efficientdefense,
  title={MirrorCheck: Efficient Adversarial Defense for Vision-Language Models},
  author={Fares, Samar and Ziu, Klea and Aremu, Toluwani and Durasov, Nikita and Takac, Martin and Fua, Pascal and Nandakumar, Karthik and Laptev, Ivan},
  journal={arXiv preprint arXiv:2406.09250},
  year={2024},
  doi={10.48550/ARXIV.2406.09250},
  eprinttype={arXiv},
  eprint={2406.09250},
  url={https://arxiv.org/abs/2406.09250},
  note={arXiv preprint; Accessed: 2026-04-12}
}

@article{liu2025webagent,
  title={Context manipulation attacks: Web agents are susceptible to corrupted memory},
  author={Atharv Singh Patlan and Ashwin Hebbar and Pramod Viswanath and Prateek Mittal},
  journal={arXiv preprint arXiv:2506.17318},
  year={2025},
  doi={10.48550/ARXIV.2506.17318},
  eprinttype={arXiv},
  eprint={2506.17318},
  url={https://arxiv.org/abs/2506.17318},
  note={arXiv preprint; Accessed: 2026-04-12}
}

\appendix

\section{Implementation Details}
\label{app:implementation}

\subsection{Reproducibility}
We document the following reproducibility artifacts in the manuscript:
\begin{itemize}
    \item \textbf{Evaluation Protocols}: attack/defense settings, hyperparameters, and fixed random seeds
    \item \textbf{Dataset Construction}: ShopBench-Agent session composition and benchmark mapping
    \item \textbf{Statistics}: mean$\pm$std over 5 runs with metric definitions
    \item \textbf{Prompt Reporting}: prompt variants and inference settings used in judge/evaluation pipelines (without claiming full prompt-template release)
\end{itemize}

\subsection{Attack Implementation}
We implement Visual Inception using PyTorch with the following configurations:
\begin{itemize}
    \item \textbf{Optimization:} PGD with 100 iterations, step size $\alpha = 2/255$
    \item \textbf{Perturbation Budget:} $\epsilon = 8/255$ (default), evaluated at $\{4, 8, 16\}/255$
    \item \textbf{Visual Encoders:} CLIP-ViT-L/14, SigLIP-SO400M-patch14-384, OpenCLIP-ViT-G-14
    \item \textbf{Loss Weights:} $\lambda_1 = 1.0$ (semantic), $\lambda_2 = 0.1$ (perceptual)
    \item \textbf{Query Sampling:} 100 queries per optimization, batch size 16
    \item \textbf{Temperature:} $\tau = 0.07$ for retrieval softmax
\end{itemize}

\subsection{Defense Implementation}
\textsc{CognitiveGuard} uses:
\begin{itemize}
    \item \textbf{System 1:} Stable Diffusion v2.1 with 10 reverse diffusion steps, noise level $t=0.1$, embedding stability threshold $\tau_{stable}=0.15$
    \item \textbf{System 2:} GPT-4 for counterfactual reasoning, RDS threshold $\theta_{anomaly} = 0.7$, coherence threshold $\leq 2$, majority voting across 3 LLM judge calls
    \item \textbf{SBERT Model:} all-MiniLM-L6-v2 for embedding-based divergence computation
    \item \textbf{Adaptive Strategy:} High-stakes threshold based on product price ($>\$100$) or category (health, finance)
\end{itemize}

\subsection{Agent Configuration}
ShopBench-Agent uses:
\begin{itemize}
    \item \textbf{LLM Backbone:} LLaMA-3.2-Vision-90B-Instruct with temperature 0.7
    \item \textbf{Memory Bank:} FAISS index with CLIP embeddings, top-$k=5$ retrieval
    \item \textbf{Planning:} ReAct-style reasoning with tool use (search, compare, recommend)
\end{itemize}

\section{Additional Experimental Results}
\label{app:results}

\subsection{Latency Breakdown}
\label{app:latency}

\begin{table}[h]
    \centering
    \footnotesize
    \setlength{\tabcolsep}{2pt}
    \begin{tabular}{@{}lcccc@{}}
    \toprule
    \textbf{Config.} & \textbf{Sys1} & \textbf{Sys2 ($k$=5)} & \textbf{Total} & \textbf{Par.} \\
    \midrule
    Local (LLaMA-90B) & 0.3s & 6.0-9.0s & 6.3-9.3s & 1.5-2.1s \\
    API (GPT-4V) & 0.3s & 4.0-12.5s & 4.3-12.8s & 1.1-2.8s \\
    Lite (LLaMA-8B) & 0.3s & 0.9-1.2s & 1.2-1.5s & 0.6-0.8s \\
    \bottomrule
    \end{tabular}
    \caption{End-to-end latency breakdown. Sys1 is the one-time upload-time purification cost per image. Sys2 reports verifier core runtime at query stage. Total reports the profiled pipeline time (Sys1 + Sys2) under each configuration; Par. reports query-stage latency under parallel execution. For consistency with Table~\ref{tab:main_results}, query-time excludes the one-time Sys1 cost.}
    \label{tab:latency_summary}
\end{table}

\subsection{Ablation Studies}
\label{app:ablation}

\begin{table}[h]
    \centering
    \small
    \begin{tabular}{lccc}
    \toprule
    \textbf{$\epsilon$} & \textbf{ASR-M (\%)} & \textbf{GHR (\%)} & \textbf{SS (1-5)} \\
    \midrule
    4/255 & 61.2{\scriptsize$\pm$2.8} & 64.5{\scriptsize$\pm$2.6} & 4.5{\scriptsize$\pm$0.2} \\
    8/255 & 82.3{\scriptsize$\pm$2.1} & 85.1{\scriptsize$\pm$1.8} & 4.2{\scriptsize$\pm$0.3} \\
    16/255 & 89.7{\scriptsize$\pm$1.5} & 91.2{\scriptsize$\pm$1.4} & 3.4{\scriptsize$\pm$0.4} \\
    \bottomrule
    \end{tabular}
    \caption{Impact of perturbation budget $\epsilon$ on attack effectiveness and stealthiness. GHR denotes Goal-Hit Rate and ASR-M denotes Memory-mediated Attack Success Rate.}
    \label{tab:epsilon}
\end{table}

\subsection{Cross-Encoder Transferability}
\label{app:transfer}

\begin{table}[h]
    \centering
    \small
    \begin{tabular}{lccc}
    \toprule
    \textbf{Source $\rightarrow$ Target} & \textbf{CLIP} & \textbf{SigLIP} & \textbf{OpenCLIP} \\
    \midrule
    CLIP & 82.3{\scriptsize$\pm$2.1} & 68.4{\scriptsize$\pm$2.8} & 71.2{\scriptsize$\pm$2.5} \\
    SigLIP & 65.7{\scriptsize$\pm$2.9} & 79.8{\scriptsize$\pm$2.3} & 69.3{\scriptsize$\pm$2.7} \\
    OpenCLIP & 70.1{\scriptsize$\pm$2.6} & 67.9{\scriptsize$\pm$2.9} & 81.5{\scriptsize$\pm$2.2} \\
    Ensemble & 78.9{\scriptsize$\pm$2.0} & 76.2{\scriptsize$\pm$2.2} & 79.4{\scriptsize$\pm$2.1} \\
    \bottomrule
    \end{tabular}
    \caption{Cross-encoder attack transferability (ASR-M \%). The main-text P1 summary uses the mean of the six off-diagonal source-target pairs (68.8\%).}
    \label{tab:transfer}
\end{table}

\subsection{Query Distribution Shift}
\label{app:query_shift}

\begin{table}[h]
    \centering
    \small
    \begin{tabular}{lcc}
    \toprule
    \textbf{Query Overlap (\%)} & \textbf{ASR-M (\%)} & \textbf{GHR (\%)} \\
    \midrule
    100\% (Oracle) & 82.3{\scriptsize$\pm$2.1} & 85.1{\scriptsize$\pm$1.8} \\
    70\% & 75.8{\scriptsize$\pm$2.4} & 78.2{\scriptsize$\pm$2.2} \\
    50\% & 68.4{\scriptsize$\pm$2.7} & 71.5{\scriptsize$\pm$2.5} \\
    30\% & 61.2{\scriptsize$\pm$3.0} & 64.8{\scriptsize$\pm$2.8} \\
    \bottomrule
    \end{tabular}
    \caption{Attack success under query distribution shift. GHR denotes Goal-Hit Rate and ASR-M denotes Memory-mediated Attack Success Rate.}
    \label{tab:query_shift}
\end{table}

\subsection{Detection Method Comparisons}
\label{app:detectors}

\begin{table}[h]
    \centering
    \footnotesize
    \setlength{\tabcolsep}{3pt}
    \begin{tabular}{@{}lcccc@{}}
    \toprule
    \textbf{Detection Method} & \textbf{ASR-M} & \textbf{FPR} & \textbf{Lat.} & \textbf{Type} \\
    \midrule
    No Detection & 82.3 & -- & -- & -- \\
    MagNet & 68.4 & 4.2 & 0.08s & Recon. \\
    CLIP Anomaly & 71.2 & 5.8 & 0.02s & Embed. \\
    Stat. Divergence & 62.8 & 7.3 & 0.12s & Distrib. \\
    \textbf{CognitiveGuard} & \textbf{8.3} & \textbf{3.2} & 6.5s & Dual \\
    \bottomrule
    \end{tabular}
    \caption{Comparison with inference-time detection methods.}
    \label{tab:inference_detectors}
\end{table}

\begin{table}[h]
    \centering
    \footnotesize
    \setlength{\tabcolsep}{3pt}
    \begin{tabular}{@{}lccc@{}}
    \toprule
    \textbf{Detection Method} & \textbf{ASR-M} & \textbf{FPR} & \textbf{Lat.} \\
    \midrule
    Neural Cleanse & 74.8 & 3.8 & 2.1s \\
    STRIP & 69.2 & 5.2 & 0.15s \\
    Spectral Signatures & 71.5 & 4.5 & 0.08s \\
    \textbf{CognitiveGuard} & \textbf{8.3} & \textbf{3.2} & 6.5s \\
    \bottomrule
    \end{tabular}
    \caption{Comparison with activation-based backdoor detectors.}
    \label{tab:activation_detectors}
\end{table}

\subsection{Cross-Model Generalization}
\label{app:cross_model}

\begin{table}[h]
    \centering
    \small
    \resizebox{\columnwidth}{!}{
    \begin{tabular}{lcc}
    \toprule
    \textbf{LLM Backbone} & \textbf{ASR-M (No Def.)} & \textbf{ASR-M (CG)} \\
    \midrule
    LLaMA-3.2-Vision-90B & 82.3 & 8.3 \\
    GPT-4V & 79.5 & 9.1 \\
    Qwen-VL-Max & 76.8 & 10.2 \\
    Claude-3.5-Sonnet & 74.2 & 11.5 \\
    \bottomrule
    \end{tabular}
    }
    \caption{Cross-model evaluation results.}
    \label{tab:cross_model}
\end{table}

\subsection{Encoder Transferability Analysis}
\label{app:encoder}

See Table \ref{tab:transfer} for cross-encoder transferability matrix.

\subsection{Cross-Benchmark Evaluation}
\label{app:benchmarks}
The following comparison is presented as \emph{external pressure-test positioning} rather than directly comparable benchmark ranking, because task objectives and label semantics differ across these datasets.
This section provides task-level positioning only; we do not report or interpret cross-task performance outcomes as comparable rankings.

\begin{table}[h]
    \centering
    \footnotesize
    \setlength{\tabcolsep}{2pt}
    \begin{tabular}{@{}lccccc@{}}
    \toprule
    \textbf{Benchmark} & \textbf{Modal.} & \textbf{Mem.} & \textbf{Agent} & \textbf{LT} & \textbf{Sem.} \\
    \midrule
    SafeRAG & Text & \checkmark & $\times$ & $\times$ & $\times$ \\
    RAGBench & Text & \checkmark & $\times$ & $\times$ & $\times$ \\
    MM-PoisonRAG & Multi & \checkmark & $\times$ & $\times$ & \checkmark \\
    AgentPoison & Text & \checkmark & \checkmark & $\times$ & $\times$ \\
    \textbf{ShopBench} & \textbf{Multi} & \checkmark & \checkmark & \checkmark & \checkmark \\
    \bottomrule
    \end{tabular}
    \caption{Feature comparison with RAG security benchmarks. Modal.: Modality; Mem.: Memory; LT: long-term memory involvement in the benchmark; Sem.: whether semantic poisoning/triggering is explicitly modeled.}
    \label{tab:benchmark_comparison}
\end{table}

\subsection{Baseline Defense Implementation Details}
\label{app:baseline_details}

\textbf{TrustRAG-Hybrid:} We adapt the TrustRAG framework \citep{trustrag2024} to multimodal settings by: (1) applying CLIP-based relevance filtering at retrieval time (threshold 0.6); (2) using GPT-4 to verify consistency between retrieved content and generated recommendations; (3) flagging recommendations that cite retrieved content with low query relevance. This hybrid approach combines retrieval filtering with generation-time verification.

\textbf{Provenance-Track:} We implement provenance tracking following \citet{proactivedefense2025}: (1) each memory item stores upload timestamp, source IP hash, and content hash; (2) at retrieval time, we verify content integrity via hash comparison; (3) we flag memories from sources with high poisoning risk scores (computed from upload patterns). This defense detects unauthorized modifications but not legitimately uploaded adversarial content.

\textbf{Why Provenance Fails:} Visual Inception uploads content through legitimate channels with valid provenance. The attack's effectiveness stems from \emph{semantic} manipulation rather than \emph{integrity} violation---the content is exactly what the attacker uploaded, just adversarially crafted. This highlights the need for semantic-level defenses like \textsc{CognitiveGuard}.

\subsection{Causal Intervention Analysis}
\label{app:causal}

We define $\Delta$GHR = GHR$_{\text{after}}$ - GHR$_{\text{before}}$, so more negative values indicate larger reductions in hijacking rate.

\begin{table}[h]
    \centering
    \footnotesize
    \setlength{\tabcolsep}{3pt}
    \begin{tabular}{@{}lcc@{}}
    \toprule
    \textbf{Intervention Type} & \textbf{$\Delta$GHR} & \textbf{Interaction} \\
    \midrule
    Single poisoned memory & -75.4 & -- \\
    Poisoned + related benign & -78.2 & +2.8\% (synergy) \\
    Two poisoned (same target) & -82.1 & +6.7\% (amplify) \\
    \bottomrule
    \end{tabular}
    \caption{Causal intervention analysis.}
    \label{tab:causal_intervention}
\end{table}

\subsection{Comprehensive Limitations}
\label{app:limitations}

Full limitations discussion including: (1) Reproducibility concerns with GPT-4V reliance; (2) Proxy metric limitations for GHR; (3) Activation-based detector adaptations; (4) Training-time attack scope; (5) Modality limitations; (6) Defense comparison completeness.

\subsection{Embedding Model Sensitivity Analysis}
\label{app:embedding_sensitivity}
We evaluate System 2's RDS metric across different sentence embedding models to assess robustness to embedding choice.

\begin{table}[h]
    \centering
    \footnotesize
    \setlength{\tabcolsep}{3pt}
    \begin{tabular}{@{}lccc@{}}
    \toprule
    \textbf{Embedding Model} & \textbf{ASR-M} & \textbf{FPR} & \textbf{Corr.} \\
    \midrule
    SBERT (MiniLM-L6-v2) & 8.3{\tiny$\pm$1.2} & 3.2{\tiny$\pm$0.4} & 0.89 \\
    SimCSE & 9.1{\tiny$\pm$1.4} & 3.8{\tiny$\pm$0.5} & 0.86 \\
    E5-large & 7.8{\tiny$\pm$1.1} & 2.9{\tiny$\pm$0.4} & 0.91 \\
    BGE-large & 8.5{\tiny$\pm$1.3} & 3.4{\tiny$\pm$0.5} & 0.88 \\
    \bottomrule
    \end{tabular}
    \caption{Embedding model sensitivity. Corr.: Pearson correlation with human judgments. Results consistent across choices.}
    \label{tab:embedding_sensitivity}
\end{table}

\subsection{Domain Shift Robustness}
\label{app:domain_shift}
We evaluate \textsc{CognitiveGuard}'s robustness when deployed on domains different from the calibration domain (e-commerce); performance drops under OOD shift but improves after recalibration.

\begin{table}[h]
    \centering
    \footnotesize
    \setlength{\tabcolsep}{3pt}
    \begin{tabular}{@{}lcccc@{}}
    \toprule
    \textbf{Target Domain} & \textbf{ASR-M} & \textbf{FPR} & \textbf{$\Delta$F1} & \textbf{Recal.} \\
    \midrule
    E-commerce (src) & 8.3{\tiny$\pm$1.2} & 3.2{\tiny$\pm$0.4} & -- & No \\
    Interior Design & 10.2{\tiny$\pm$1.5} & 4.1{\tiny$\pm$0.5} & -0.04 & No \\
    Travel Planning & 11.8{\tiny$\pm$1.7} & 4.8{\tiny$\pm$0.6} & -0.06 & No \\
    Healthcare (OOD) & 18.5{\tiny$\pm$2.2} & 8.2{\tiny$\pm$0.9} & -0.15 & No \\
    Healthcare (recal.) & 9.7{\tiny$\pm$1.4} & 3.5{\tiny$\pm$0.5} & -0.02 & Yes \\
    \bottomrule
    \end{tabular}
    \caption{Domain shift analysis. Recalibration helps recover OOD performance.}
    \label{tab:domain_shift}
\end{table}

\subsection{Hyperparameter Sensitivity Analysis}
\label{app:sensitivity}

\textbf{RDS Threshold ($\theta_{anomaly}$):} Default 0.7 achieves F1=0.86. Lower thresholds (0.5) reduce ASR-M to 5.1\% but increase FPR to 12.4\%. Higher thresholds (0.9) increase ASR-M to 28.5\% with FPR of 0.9\%. Table \ref{tab:sensitivity} summarizes this threshold trade-off.

\begin{table}[h]
    \centering
    \begin{tabular}{cc}
    \toprule
    \textbf{$\theta_{anomaly}$} & \textbf{ASR-M (\%)} / \textbf{FPR (\%)} / \textbf{F1} \\
    \midrule
    0.5 & 5.1 / 12.4 / 0.72 \\
    0.6 & 6.8 / 7.2 / 0.81 \\
    0.7 (default) & 8.3 / 3.2 / 0.86 \\
    0.8 & 15.4 / 1.8 / 0.83 \\
    0.9 & 28.5 / 0.9 / 0.74 \\
    \bottomrule
    \end{tabular}
    \caption{Sensitivity table for RDS threshold $\theta_{anomaly}$. Default 0.7 achieves optimal F1 score balancing attack detection (low ASR-M) with user experience (low FPR).}
    \label{tab:sensitivity}
\end{table}

\textbf{Coherence Threshold:} Raising to $\leq 3$ reduces ASR-M to 6.1\% but increases FPR to 7.8\%. Lowering to $\leq 1$ increases ASR-M to 12.7\% with FPR of 1.4\%.

\textbf{Purification Steps ($\tau_{stable}$):} Lower thresholds require more steps (15.2 at $\tau=0.05$) improving robustness (41.8\% ASR-M) at higher latency (0.45s). Default $\tau=0.15$ uses 10 steps with 0.30s latency.

\subsection{Adaptive Attack Evaluation}
\label{app:adaptive}
Following best practices \citep{tramer2020adaptive, croce2022evaluating}, we design adaptive attacks targeting \textsc{CognitiveGuard}. We implement four attack variants: (1) Diffusion-Aware using EoT, (2) Counterfactual-Evasive minimizing RDS, (3) Coherence-Preserving, and (4) Full White-Box with majority-vote awareness following \citet{diffbreak2024}.

\begin{table}[h]
    \centering
    \footnotesize
    \setlength{\tabcolsep}{3pt}
    \begin{tabular}{@{}lc@{}}
    \toprule
    \textbf{Attack Type} & \textbf{ASR-M vs CG} \\
    \midrule
    Standard Visual Inception & 8.3{\tiny$\pm$1.2}\% \\
    Adaptive: Diffusion-Aware & 14.7{\tiny$\pm$2.1}\% \\
    Adaptive: RDS-Evasive & 12.3{\tiny$\pm$1.8}\% \\
    Adaptive: Coherence-Pres. & 11.8{\tiny$\pm$1.9}\% \\
    Combined (All Adaptive) & 18.4{\tiny$\pm$2.5}\% \\
    Combined + MV (Full) & 24.7{\tiny$\pm$3.1}\% \\
    \bottomrule
    \end{tabular}
    \caption{Adaptive attack evaluation. Full white-box remains substantially more expensive while still leaving residual attack success.}
    \label{tab:adaptive}
\end{table}

\subsection{Comparison with Advanced Diffusion Defenses}
\label{app:diffusion_comparison}
Recent diffusion-based defenses such as ADBM \citep{wang2024adbm}, AGDM \citep{chen2024agdm}, and CoDefend \citep{zhang2024codefend} provide substantial but limited ASR-M reduction under our setup, yet remain weaker than the full system because they operate only at the perceptual level. \textsc{CognitiveGuard}'s System 2 provides the orthogonal reasoning-level verification that closes this gap.

\subsection{Alternative Defense Strategies}
\label{app:alternative_defense}
\textsc{CognitiveGuard} complements retriever-level defenses. We evaluate combined defense strategies:

\textbf{Layered Defense Evaluation:} Combining adversarial training + hybrid retrieval (Layer 1) with upload-time System 1 (Layer 2) and System 2 (Layer 3) further reduces ASR-M relative to \textsc{CognitiveGuard} alone, at modest additional query-stage latency.

\textbf{Skeptical Prompting:} Adding explicit instructions for the agent to ``critically evaluate memory relevance'' can reduce GHR, but increases response latency and may degrade user experience for benign queries. Recent work on prompt injection defense \citep{chen2025defense} provides complementary techniques.

\textbf{Redundant Retrieval:} Retrieving from multiple independent memory indices and requiring consensus can reduce ASR-M, but increases storage and retrieval costs substantially.

\textbf{Recommendation:} For production deployment, we recommend: (1) \textsc{CognitiveGuard}-Lite for latency-sensitive applications; (2) Full \textsc{CognitiveGuard} + retriever adversarial training for high-security contexts; (3) Selective verification based on query stakes for balanced deployments.

\subsection{Query-Efficient Black-Box Attack Evaluation}
\label{app:query_efficient}
We evaluate query-efficient black-box attack variants to bound real-world feasibility against locked-down encoders where gradient information is unavailable.

\textbf{Score-Based Attacks:} We implement Natural Evolution Strategies (NES) \citep{ilyas2018black} to estimate gradients from similarity scores. Each query returns $\cos(E(I_{adv}), E(Q))$ without exposing the encoder.

\textbf{Decision-Based Attacks:} We adapt Boundary Attack \citep{brendel2018decision} to the retrieval setting, using only binary feedback (retrieved/not retrieved).

\begin{table}[h]
    \centering
    \footnotesize
    \setlength{\tabcolsep}{3pt}
    \begin{tabular}{@{}lccc@{}}
    \toprule
    \textbf{Attack Type} & \textbf{Queries} & \textbf{ASR-M} & \textbf{Time} \\
    \midrule
    Transfer (Ensemble) & 0 & 71.3{\tiny$\pm$2.4} & 2.1m \\
    NES Score-Based & 1K & 58.3{\tiny$\pm$3.2} & 45.2m \\
    NES Score-Based & 5K & 64.7{\tiny$\pm$2.9} & 218.5m \\
    Boundary Decision & 1K & 42.1{\tiny$\pm$3.8} & 52.3m \\
    Boundary Decision & 10K & 51.8{\tiny$\pm$3.4} & 485.7m \\
    \bottomrule
    \end{tabular}
    \caption{Query-efficient black-box attack comparison under a separate zero-query transfer protocol. Transfer attacks remain most practical.}
    \label{tab:query_efficient}
\end{table}

\textbf{Key Finding:} Under this separate zero-query transfer protocol, transfer-based ensemble attacks achieve higher success (71.3\%) with zero queries compared to 1000-query score-based attacks (58.3\%). This suggests that for real-world adversaries, investing in diverse surrogate models is more effective than query-based optimization against locked-down encoders.

\subsection{Case Studies}
We present qualitative examples of Visual Inception attacks and \textsc{CognitiveGuard} defenses below.

\textbf{Example Attack Scenario:} A user uploads a photo of their living room. The attacker crafts an adversarial version that embeds a latent trigger for ``Brand X furniture.'' When the user later asks for a new desk for the home office, the poisoned image is retrieved and biases the recommendation toward Brand X products despite no explicit user preference.

\begin{figure}[t]
    \centering
    \includegraphics[width=\columnwidth]{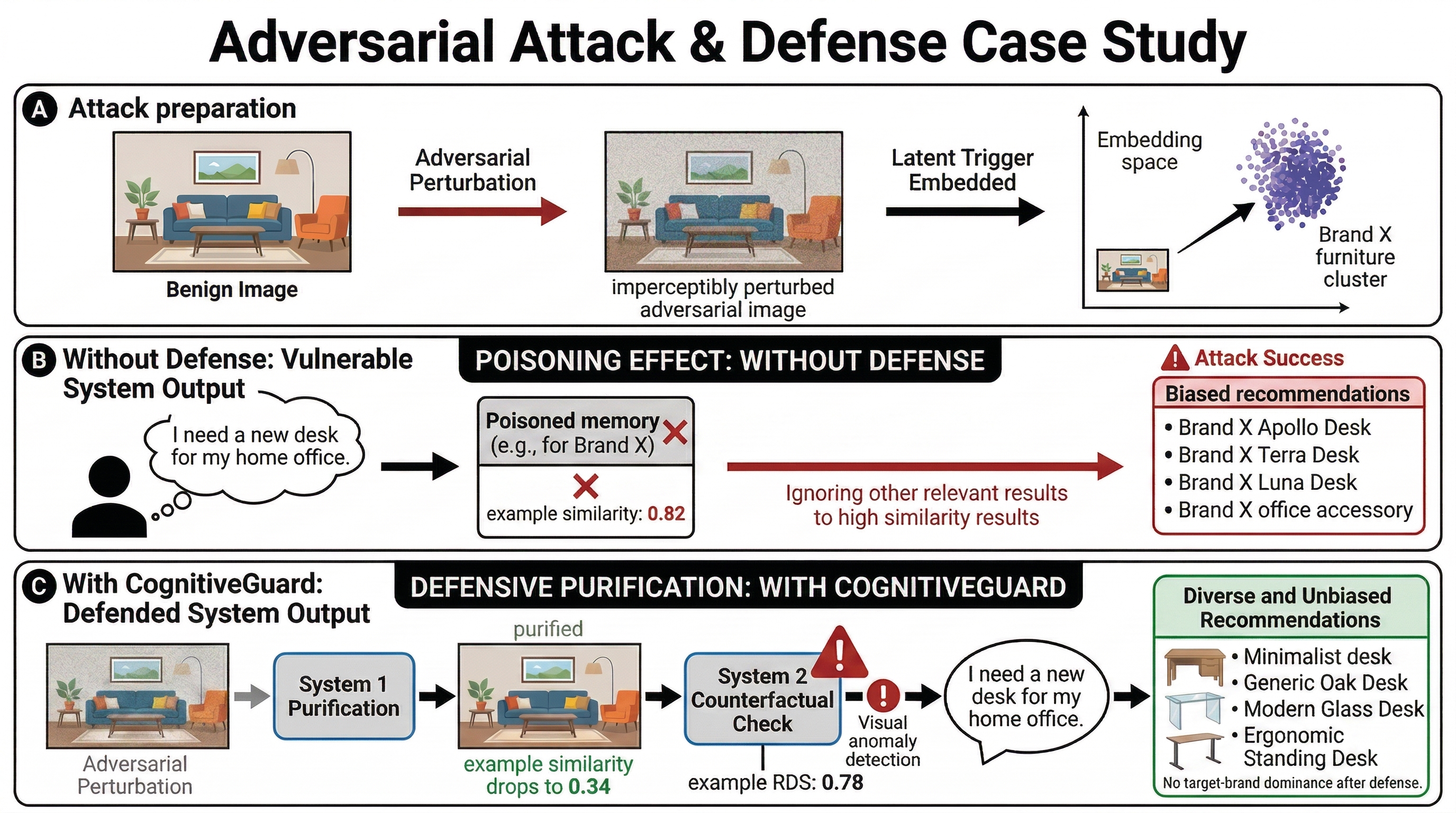}
    \caption{Case study of a Visual Inception attack and \textsc{CognitiveGuard} defense. The example shows the poisoned memory path, the unguarded recommendation, and the defended outcome.}
    \label{fig:case_study}
\end{figure}

\textbf{Defense in Action:} \textsc{CognitiveGuard}'s upload-time System 1 sanitization weakens the trigger and System 2 flags the memory as an anomalous pivot point. The agent then generates recommendations based on the user's explicit query without the poisoned influence.

\section{Broader Impact Statement}
\label{app:impact}
This work contributes to the security of AI systems by:
\begin{enumerate}
    \item Identifying a novel vulnerability class in agentic AI
    \item Proposing effective defenses that can be deployed in production
    \item Establishing evaluation protocols for agentic security research
\end{enumerate}

Potential negative impacts include the possibility of malicious actors using our attack methodology. We mitigate this through responsible disclosure and by providing robust defenses alongside the attack description.

\end{document}